\documentclass[10pt,twocolumn,letterpaper]{article}

\usepackage{cvpr}      

\usepackage{graphicx}
\usepackage{amsmath}
\usepackage{amssymb}
\usepackage{booktabs}
\usepackage{float}
\usepackage{multirow}
\usepackage{dsfont}
\usepackage{subfloat}
\graphicspath{ {./images/} }
\usepackage[pagebackref,breaklinks,colorlinks]{hyperref}
\usepackage{enumitem}

\hypersetup{
    colorlinks=true,
    citecolor=blue,
    linkcolor=blue,
    filecolor=magenta,      
    urlcolor=blue,
}
\usepackage[capitalize]{cleveref}
\crefname{section}{Sec.}{Secs.}
\Crefname{section}{Section}{Sections}
\Crefname{table}{Table}{Tables}
\crefname{table}{Tab.}{Tabs.}

\begin{document}

\title{Full Reference Video Quality Assessment for Machine Learning-Based Video Codecs}

\author{Abrar Majeedi\\
University of Wisconsin-Madison\\
{\tt\small majeedi@wisc.edu}
\and
Babak Naderi\\
Microsoft\\
{\tt\small babaknaderi@microsoft.com}
\and
Yasaman Hosseinkashi\\
Microsoft\\
{\tt\small yahossei@microsoft.com}
\and
Juhee Cho\\
Microsoft\\
{\tt\small juhcho@microsoft.com}
\and
Ruben Alvarez Martinez\\
Microsoft\\
{\tt\small rubenal@microsoft.com}
\and
Ross Cutler\\
Microsoft\\
{\tt\small ross.cutler@microsoft.com}
}

\maketitle

\begin{abstract}
Machine learning-based video codecs have made significant progress in the past few years. A critical area in the development of ML-based video codecs is an accurate evaluation metric that does not require an expensive and slow subjective test. We show that existing evaluation metrics that were designed and trained on DSP-based video codecs are not highly correlated to subjective opinion when used with ML video codecs due to the video artifacts being quite different between ML and video codecs. We provide a new dataset of ML video codec videos that have been accurately labeled for quality. We also propose a new full reference video quality assessment (FRVQA) model that achieves a Pearson Correlation Coefficient (PCC) of 0.99 and a Spearman's Rank Correlation Coefficient (SRCC) of 0.99 at the model level. We make the dataset and FRVQA model open source to help accelerate research in ML video codecs, and so that others can further improve the FRVQA model.
\end{abstract}

\section{Introduction}
\label{sec:intro}

Internet traffic statistics show that internet video traffic will be 82\% of all consumer Internet traffic by 2022, up from 73\% in 2017, which is a compound annual growth rate of 34\%\cite{VNICompleteForecast2018}. This number is only expected to grow, with the ever-increasing popularity of the video social media platforms like TikTok, Reels, and YouTube, and videoconferencing applications like Microsoft Teams and Zoom. Since the physical internet infrastructure is limited and cannot be scaled up fast enough to keep up with the exponential growth of internet traffic, video transmission has the potential to choke up the internet\cite{hechtBandwidthBottleneckThat2016}. Video compression technologies enable video to be streamed across the internet at a small fraction of the uncompressed bandwidth, which enables video streaming and video conferencing applications to be possible. These video compression methods involve fast and efficient hand-coded digital signal processing (DSP) algorithms to reduce the size of the video files by over 1000x with an acceptable reduction in quality. 

In the late 1980s, the common H.26X video compression standards were introduced, starting with H.261 in 1988 \cite{h261_video_1998}, followed by H.262 in 1994 \cite{noauthor_isoiec_1995}, H.263 in 1996 \cite{rijkse_h263_1996}, H.264 in 2003 \cite{wiegand_overview_2003}, H.265 in 2013 \cite{sullivan_overview_2012}, and the latest one H.266 in 2020 \cite{bross_overview_2021}. Each one of these DSP codecs roughly doubled the coding efficiency of their predecessors, and the average time between versions for the past three codecs is 8.5 years.
\begin{figure}
    \centering
  \includegraphics[width = 3.2in]{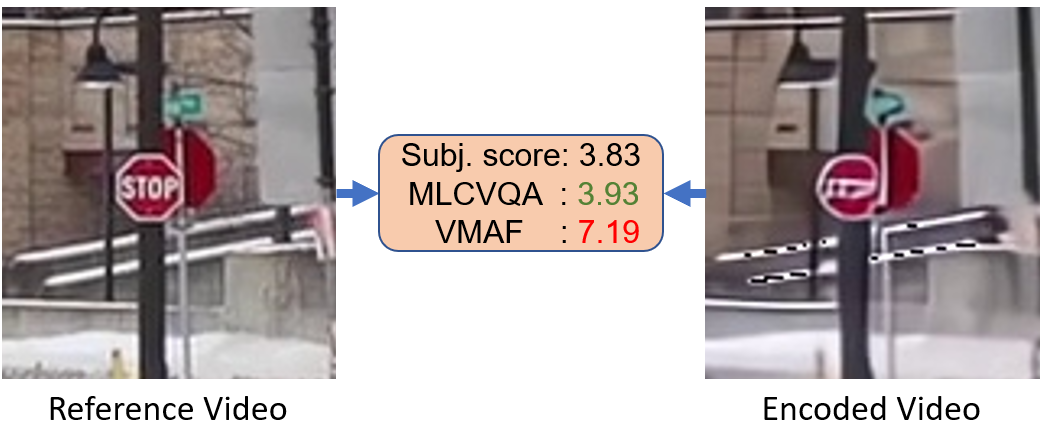}
  \caption{Full reference video quality assessment. The image on the left is from the original video, while the image on the right is from the compressed version. The scores are in the range of 1-9, where 9 is the best.}
  \label{fig:solo_demo}
\end{figure}

Machine learning (ML) is now actively being used to improve the coding efficiency of video codecs. 
However, the vast majority of ML video codecs 
\cite{liDeepContextualVideo2021,liHybridSpatialTemporalEntropy2022, luDVCEndToEndDeep2019, luEndtoEndLearningFramework2021,lin_m-lvc_2020,agustsson_scale-space_2020} only evaluate the codec quality using metrics like Peak Signal-to-Noise ratio (PSNR) \cite{gonzalezDigitalImageProcessing2006} and Multi-Scale Structural Similarity Index Measure (MS-SSIM) \cite{wangMultiscaleStructuralSimilarity2003a}, which as we show in the paper, are poorly correlated with human subjective quality assessment.

The development and adoption of ML video codecs have faced a roadblock in terms of perceptual quality evaluation, i.e., measuring how good or bad a human viewer perceives the videos compressed by an ML video codec. ML video codecs have different artifacts than DSP-based video codecs, as shown in Fig.~\ref{fig:solo_demo}. Existing FRVQAs do not work well with ML video codecs (see Section \ref{sec:results}). 
The gold standard approach for measuring perceptual quality is subjective testing in which a group of test participants watch the processed videos and rate their quality following different standardized test methods depending on the use case, e.g., ITU-T Rec.~P.910 \cite{itu-t_recommendation_p910_subjective_2021}, P.911 \cite{itu-t_recommendation_p911_subjective_1998}, P.913 \cite{itu-t_recommendation_p913_methods_2021}, and ITU-R BT.500 \cite{itu-r_recommendation_bt500_methodologies_2019}. Besides the test method, the standards define the viewing conditions, participant qualification tests, and device properties to reduce their effects on subjective ratings.
The aggregated ratings represent the processing system's perceptual quality and are used to directly compare the performance of different codecs.
This process is costly and time-consuming, especially since we expect many versions of a codec during development, differing in model architectures, model sizes, and hyper-parameters among other possible design choices.

In order to address this issue, we propose an ML-based video quality assessment model that achieves a PCC of 0.99, SRCC of 0.99, and Kendall's Tau-b 95 (see Section \ref{sec:metrics}) of 0.93 with the subjective scores, outperforming the existing state-of-the-art objective metrics. We also release a dataset with a wide variety of videos encoded by many ML video codecs, complete with subjective scores for each video-codec pair. Our model can be used to get accurate perceptual quality measures for videos, and thus allow for faster development and evaluation of ML video codecs. We make our code, trained model, and the full dataset available to the community to accelerate the development and evaluation of ML video codecs\footnote{\url{https://github/microsoft/MLCVQA}}.

Our contributions are:
\begin{enumerate}[itemsep=2mm, parsep=0pt] 
    \item We provide the first FRVQA for ML video codecs and show it provides best-in-class results.
    \item We provide the first large dataset, MLVC-FRVQA, for training and testing FRVQA models for ML video codecs.
    \item We provide a new statistical metric, Tau-b 95, for evaluating perceptual metrics and provide the theoretical limit to an FRVQA on the given test set with this metric.
\end{enumerate}

In Section \ref{sec:related} we discuss related work, in Section \ref{sec:datasets} discuss a new dataset developed for this task, in Section \ref{sec:method} describe the model and training, in Section \ref{sec:results} give the experimental results, and in Section \ref{sec:conclusion} discuss conclusions and future work.

\section{Related work}
\label{sec:related}
While learning-based methods have taken over in a majority of vision-based problem domains, video quality assessment still remains dominated by quantitative formula-based approaches. Almost all of the current video quality assessment tasks utilize metrics such as PSNR \cite{gonzalezDigitalImageProcessing2006}, Structural Similarity Index measure (SSIM) \cite{wangImageQualityAssessment2004a}, and its advanced version, MS-SSIM \cite{wangMultiscaleStructuralSimilarity2003a}. These metrics are calculated at the frame level and use pooling, usually just a mean of frame level scores, to get the final video quality score.

Netflix developed a machine learning solution, Video Multi-method Assessment Fusion (VMAF) \cite{topiwala_vmaf_2021,liVMAFJourneyContinues2018}, which predicts the perceptual quality using different objective quality metrics. Specifically, VMAF fuses two image quality metrics, Visual Information Fidelity (VIF) \cite{vuViS3AlgorithmVideo2014} and Detail Loss Metric (DLM) \cite{liImageQualityAssessment2011}. It also uses a very simple motion metric which is a pixel-wise temporal difference of the luminance channels between adjacent frames. VMAF fuses these metrics using support vector regression to calculate a weighted average, which measures the final perceptual quality score. Since VMAF relies on objective metrics and performs a frame-level temporal pooling of quality, it is not able to take into account the motion component of the video, which is more complex than pixel-wise differences in adjacent frames. As a result, it has a lower performance on videos with more motion. This limitation is discussed in detail in Section \ref{sec:results}. 

Recently, multiple end-to-end deep learning approaches have been proposed for video quality assessment. Xu et al.~proposed C3DVQA \cite{xu_c3dvqa_2020}, which consists of using 3D convolutions to account for the temporal nature of the video data. This allows the model to learn spatial and temporal variations, thus resulting in a better measurement of perceptual quality. One limitation is that it only uses the luminance channel, which hinders its performance since video codecs can include color distortions. It also downsizes the input video, which makes the subtle distortions caused by ML codecs hard to perceive.

 Sun et al.~proposed CompressedVQA \cite{sunDeepLearningBased2022} which is 
 also an end-to-end trainable method for full as well as no reference video quality assessment. Their framework consists of a feature extraction module to extract features from different layers of a convolutional neural network which are transformed by hand-coded functions into ``texture" and ``structure" features. These transformed features
are used to regress the frame level quality scores followed by a subjectively-inspired temporal
pooling strategy proposed in \cite{liQualityAssessmentIntheWild2019} which considers the memory effect of previous frames and the hysteresis effect of the next frames to the current frame. This approach does not use 3D convolutions or any learnable pooling to better model the temporal nature of the video data. Also, like C3DVQA, CompressedVQA also downsizes the input videos. 

With the success of transfer learning in computer vision, there has been considerable interest in studying the nature and re-usability of features extracted by deep models trained on large datasets such as ImageNet \cite{dengImageNetLargescaleHierarchical2009a}, and Kinetics \cite{carreiraQuoVadisAction2017}. Zhang et al.~\cite{zhangUnreasonableEffectivenessDeep2018} studied the use of deep features for perceptual quality measurement of images, and the proposed Learned Perceptual Image Patch Similarity (LPIPS) metric. Tariq et al.~\cite{tariqWhyAreDeep2020} discussed the reasoning behind why deep features work as good perceptual metrics. Li et al.~\cite{li_blindly_2022} demonstrate this idea for no-reference video quality assessment by using only the Fast pathway of the SlowFast model. In this paper, we show the effectiveness of the pre-trained SlowFast\cite{feichtenhoferSlowFastNetworksVideo2019} model (using both slow and fast pathways) at full-reference video quality assessment. 

\section{Datasets}
\label{sec:datasets}
\subsection{CLIC challenge dataset}
\label{sec:clic_dataset}

The CVPR 2022 CLIC video compression challenge\footnote{\url{http://compression.cc}} resulted in 12 teams compressing 30 videos in the test set targeting 1 Mbps and 0.1 Mbps average bitrate. This included 27 combinations of codecs and bitrates, including the original sequences and two sets encoded by H.264 and AV1 \cite{han_technical_2021}. Overall, the resulting dataset contains 810 videos which are either processed by 27 encoder-bitrates combinations or are uncompressed videos. 

The Microsoft P.910 Crowdsourcing Toolkit \cite{naderi_crowdsourcing_2022} was used to assess the quality of the videos using three different subjective test methodologies. The toolkit provides an open-source implementation of ITU-T Rec.~P.910 \cite{itu-t_recommendation_p910_subjective_2021} for crowdsourcing. The CLIC 2022 dataset is available at \url{http://compression.cc}.

The Absolute Category Rating (ACR) is a single-stimulus subjective test method, in which participants watch a sequence and rate the quality on a five-point discrete scale. The Degradation Category Rating (DCR) and Comparison Category Rating (CCR) are double-stimulus test methods, in which participants watch the source and the processed clips before casting their votes. In DCR, participants are informed which sequence is the reference clip and rate the level of degradation in the processed sequence compared to the reference. However, in CCR participants are not aware of which sequence is processed and rate the quality of the second clip compared to the first clip they watched. As the source clips were included in the dataset, ACR with Hidden Reference (ACR-HR) was computed as well (see \cite{itu-t_recommendation_p910_subjective_2021} for further description of the test methods). 
On average, 21.6 valid votes per sequence in the ACR test, 12.1 in the DCR test (on a 9-point discrete scale), and 15.2 in the CCR test were collected. Table \ref{tab:test_methods} reports the correlation between Mean Opinion Scores (MOS) of different test methods aggregated in clip or model level. The highest correlations are observed within double-stimulus methods.

\begin{table}[htbp]
    \caption{Pearson correlations (upper-triangle) and Spearman's correlations (lower-triangle) between ratings collected via different test methods, aggregated in model and clip level.}
    
    \label{tab:test_methods} 
    \begin{center}
    \resizebox{\columnwidth}{!}{%
        \begin{tabular}{ c  c c c c  c   c c cc}
        \toprule
         \textbf{Test}&	 \multicolumn{4}{c}{\textbf{Clip level}}& &  \multicolumn{4}{c}{\textbf{Model level}}\\
        
        \textbf{method} & {\small \textbf{ACR}} & {\small \textbf{ACR-HR}}&	{\small\textbf{DCR}}&	{\small\textbf{CCR}} & & {\small \textbf{ACR}}& {\small \textbf{ACR-HR}} &	{\small\textbf{DCR}}&	{\small\textbf{CCR}} \\ 
        \midrule
            {\small\textbf{ACR}}&  -     & 0.975 &    0.948 & 0.920 & &   -  & 0.999  & 0.981 & 0.963 \\
            
            {\small\textbf{ACR-HR}} &  0.964 & -     &    0.939 & 0.912 &&  0.999 &  -    & 0.983 & 0.966 \\
            
            {\small\textbf{DCR}} & 0.948 & 0.935  &   -      & 0.962 && 0.976 & 0.981 &  -    & 0.990 \\
            {\small\textbf{CCR}} & 0.936 &0.928  & 0.960    &   -   && 0.987 & 0.989& 0.981  &  -    \\
        \bottomrule
        \end{tabular}
    }
    \end{center}
\end{table}

According to the P.910 recommendation, the double-stimulus test methods with an explicit reference should be employed when high-quality systems are under test, or the transparency or fidelity of the underlying process or transmission should be evaluated. These test methods lead to longer test sessions and consequently higher costs compared to the single-stimulus approach (about $\times2$). We also observed a large difference in the quality rating of clips according to single- or double-stimulus tests when they contain (1) imperfect source sequence (e.g., low image quality, or slow motioned), (2) specific impairment due to encoding process (e.g., color changes or blur effect on non-saliency area), (3) source sequence with bokeh.

As FRVQA models use both reference and processed videos for prediction, we decided to use scores from a double-stimulus subjective test method. Consequently, participants are also exposed to both reference and processed videos before casting their votes. We chose DCR scores (known as DMOS) since the test set includes only high-quality video clips and the current codecs do not improve the original video quality. We repeated the DCR subjective test (9-point scale) again to increase the number of valid votes. The results of both subjective tests highly correlated (see Table~\ref{tab:dcr_run1vs2}) leading to 26.6 valid votes per clip on average. Reported statistical metrics also indicate the presence of uncertainty in subjective ratings, even in test-retest experiments, and the importance of considering this uncertainty when evaluating objective models.
\begin{table}[htbp]
    \caption{Statistical metrics between two runs of DCR subjective tests of the CLIC 2022 dataset.}
    \small
    \label{tab:dcr_run1vs2} 
    \centering
    \resizebox{0.7\columnwidth}{!}{%
        \begin{tabular}{c c  c c c c  c }
        \toprule
         \textbf{Level} & \textbf{SRCC} & \textbf{PCC} &\textbf{Tau-b 95} &\textbf{RMSE}  \\
        
        \midrule

            Clip & 0.975 & 0.983  & 0.900 & 0.489 \\
            Model & 0.999 & 0.998 & 0.987 & 0.181\\
        \bottomrule
        \end{tabular}
    }
\end{table}

\subsection{ML Video Codec Full Reference Video Quality Assessment (MLVC-FRVQA) Dataset}
Two issues with the CVPR 2022 CLIC dataset are (1) It does not have broad coverage of spatial and temporal information (see Figure \ref{fig:dataset:siti}), and (2) It only includes high-quality videos, which is not the case in user-generated content and video conferencing scenarios. To address these issues, we created a more comprehensive dataset based on several datasets described in Table \ref{tab:dataset_sources}. Each clip in the MLVC-FRVQA is resized to the resolutions given in Table \ref{tab:dataset_resolutions} and processed by ML video codecs at the target bitrates given in Table \ref{tab:dataset_resolutions}. The MLVC-FRVQA dataset uses 7 open source ML video codec implementations \cite{liDeepContextualVideo2021,liHybridSpatialTemporalEntropy2022,hu_fvc_2021,agustsson_scale-space_2020,mentzer_vct_2022,lin_m-lvc_2020,yang_learning_2021}, 4 traditional DSP codecs (H.264/AVC, H.265/HEVC, H.266/VVC, AV1), and 2 from CVPR CLIC 2022 challengers, for a total of 13 codecs. The total number of rated clips in the MLVC-FRVQA dataset is $147\times13\times12=22,932$.

The video clips were subjectively labeled using P.910 CCR with an average number of raters of $N=25$. We used the CCR test method as a portion of videos in the MLVC-FRVQA dataset have less than perfect quality, and some of the codecs enhance the perceived quality. The MLVC-FRVQA dataset is available at \url{https://github.com/microsoft/MLCVQA}. MLVC-FRVQA is not used in the MLCVQA model in this paper due to the timing of the dataset completion. 

\begin{table}
    \centering
    \resizebox{0.95\columnwidth}{!}{%
    \begin{tabular}{c c c}
    \toprule
        \textbf{Source} & \textbf{Source clips} & \textbf{Encoded clips}\\
        \midrule
        UVG dataset  \cite{mercat_uvg_2020} & 16 & 2496 \\
        MCL-JCV \cite{wang_mcl-jcv_2016} & 30 & 4680 \\
        HEVC  \cite{bossen_common_2013} & 25 & 3900 \\
        CLIC 2022 challenge test set & 30 & 4680 \\
        MS-Webcam dataset & 46 & 7176 \\ 
        \bottomrule
        Total & \textbf{147} & \textbf{22,932}\\
        \bottomrule
    \end{tabular}
    }
    \caption{MLVC-FRVQA dataset: 147 source clips and 22,932 encoded clips}
    \label{tab:dataset_sources}
\end{table}

\begin{table}
    \centering
    \small
    \begin{tabular}{l l }
    \toprule
        \textbf{Resolution} &  \textbf{Bitrates (Kbps)} \\
        \midrule
        1080p &       1000, 500, 250, 125 \\
        720p  &       500, 250, 125, 62.5 \\
        480p  &       250, 125, 62.5, 31.25 \\
        \bottomrule
    \end{tabular}
    \caption{Resolutions and bitrates for the MLVC-FRVQA dataset}
    \label{tab:dataset_resolutions}
\end{table}

\subsection{User bias correction}
\label{sec:bias}
\begin{figure}
  \centering
    \begin{subfigure}[b]{0.7\columnwidth}
        \includegraphics[width=\columnwidth]{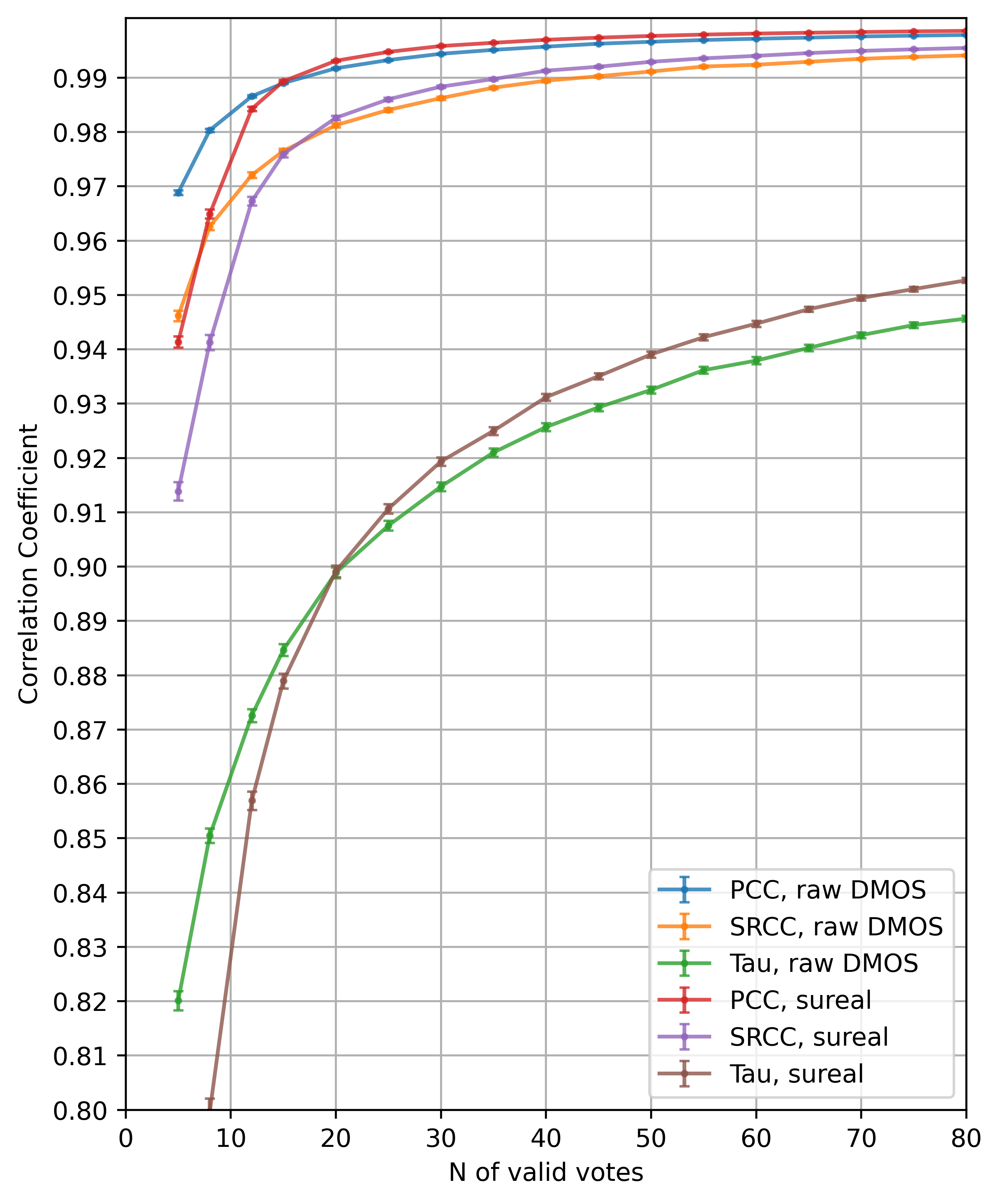}
        \caption{Changes of PCC, SRCC, and Tau}
        \label{fig:bstrp:corr}
    \end{subfigure}   
    \begin{subfigure}[b]{0.7\columnwidth}
        \includegraphics[width=\columnwidth]{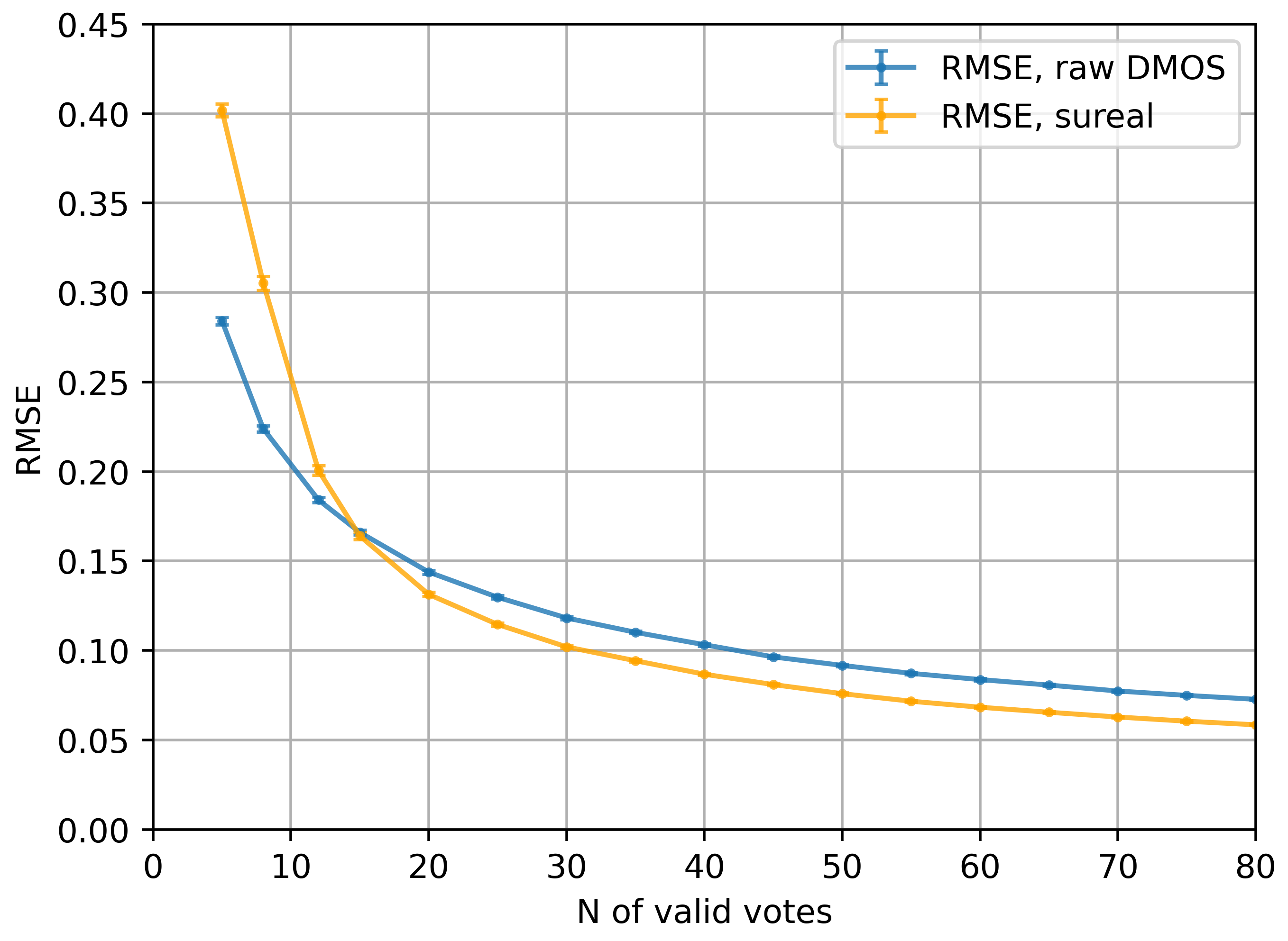}
        \caption{Changes of RMSE}
        \label{fig:bstrp:rmse}
    \end{subfigure}
  \caption{Changes of PCC, SRCC, Tau, and RMSE by increasing the number of votes per clip used for calculating raw DMOS and SUREAL corrected DMOS. Error bars around dots represent 95\% CI during 250 simulation runs.}
  \label{fig:bootstrapping_sureal_vs_rawmos}
\end{figure}

Raw measurements of subjective quality scores are often very noisy \cite{noauthor_itu-t_2020}. To get more accurate subjective quality scores and help machine learning-based FRVQA models learn easier we applied SUREAL \cite{li_recover_2017}. We first checked the effectiveness of SUREAL by performing bootstrapping simulations with the VQEG HDTV datasets and the subjective quality scores collected in \cite{naderi_crowdsourcing_2022}. This dataset contains 81 votes per clip on average and we performed similar bootstrapping simulations as in Section 4.6 in the paper. N votes were randomly selected per clip with a replacement while making sure a minimum of N or 10 votes are selected from each voter. The selected votes were used to calculate both raw DMOS and SUREAL corrected DMOS and then resulting DMOS values were compared against the ground truth DMOS values - the average of all votes collected per clip - in terms of correlations and RMSE. We repeated this process 250 times per number of selected votes (N). Figure \ref{fig:bootstrapping_sureal_vs_rawmos} shows the results. Raw DMOS aligns better with the ground truth DMOS than SUREAL corrected DMOS until the number of votes reaches around 20. The pattern is reversed after 20 votes. This may be because SUREAL is a Maximum Likelihood Estimator and it requires a certain number of samples to perform reliably per the Central Limit Theorem. Our dataset has 26 votes per clip on average and it lies on the borderline of getting benefits by applying SUREAL, so we compared our results with and without SUREAL-corrected DMOS. We trained and tested our models without any augmentation using 5-fold cross-validation. Table \ref{tab:raw_vs_sureal} summarizes the results. We observe that SUREAL does not improve our results significantly. We believe that this is mainly because the P.910 Crowdsourcing Toolkit \cite{naderi_crowdsourcing_2022} we used for data collection already applies intensive rater, environment, and device qualification tests, and includes rating quality tests, which altogether leads to less bias and noise in the aggregated data.

\begin{table}
    \caption{Results of the MLVQA model training with raw DMOS and SUREAL corrected DMOS.}
    
    \label{tab:raw_vs_sureal} 
    \begin{center}
    \resizebox{0.7\columnwidth}{!}{%
        \begin{tabular}{ c  c c c  c c}
        \toprule
          &	 \multicolumn{2}{c}{\textbf{Clip level}}& &  \multicolumn{2}{c}{\textbf{Model level}}\\
        
         \textbf{DMOS}& {\small \textbf{raw}} & {\small \textbf{SUREAL}}& & {\small \textbf{raw}}& {\small \textbf{SUREAL}} \\ 
        \midrule
            {\small\textbf{PCC}} & 0.89 & 0.89 & & 0.99 & 0.99\\
            
            {\small\textbf{SRCC}} & 0.88 & 0.89 & & 0.98 & 0.98\\
            {\small\textbf{RMSE}} & 1.18 & 1.18 & & 0.52 & 0.56\\
            {\small\textbf{Tau-b 95}} & 0.74 & 0.74 & & 0.92 & 0.92\\
            
            
        \bottomrule
        \end{tabular}
    }
    \end{center}
\end{table} 

\subsection{Spatial and temporal information distribution of MLVC-FRVQA}
\begin{figure}
    \centering
  \includegraphics[width = 1\columnwidth]{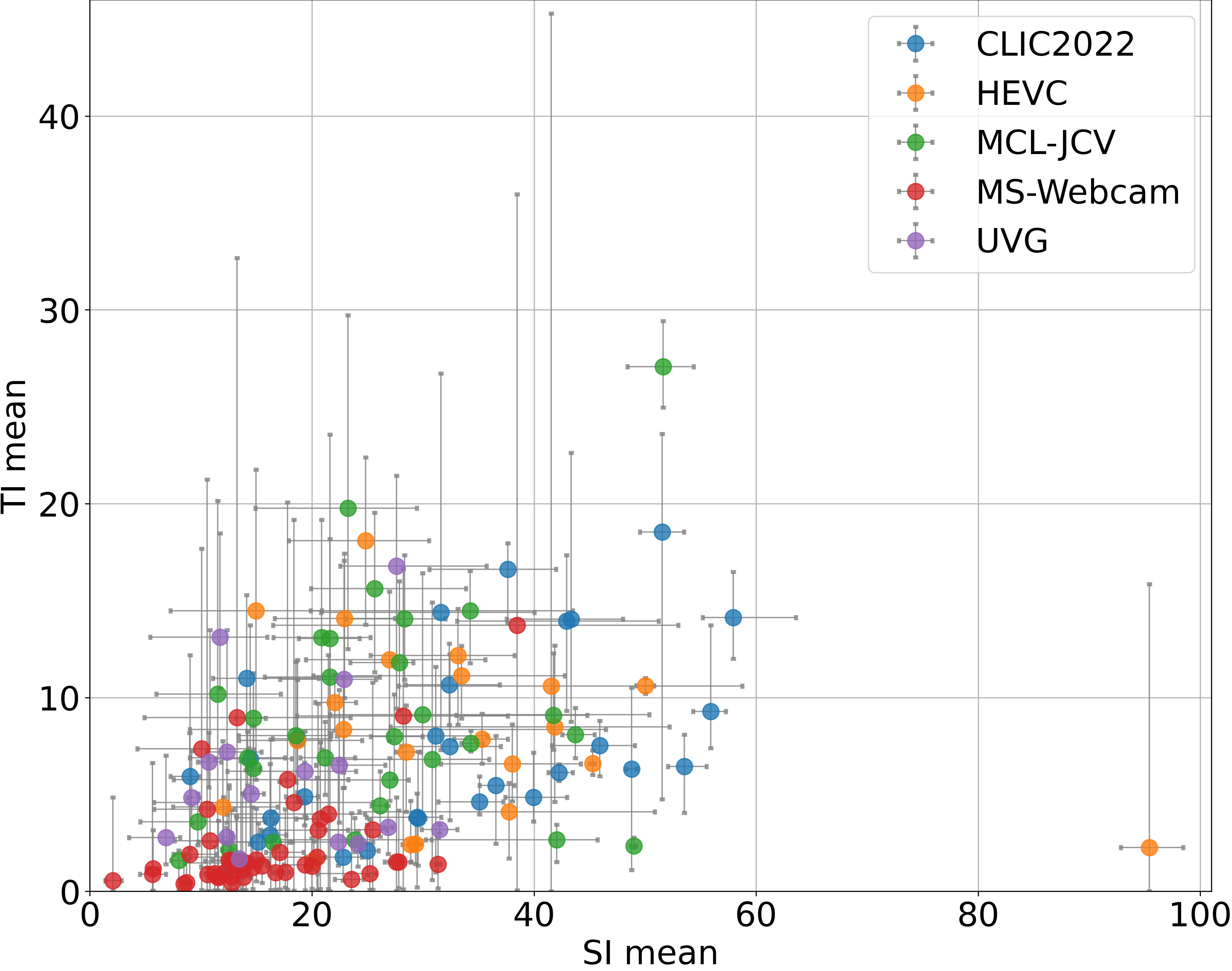}
  \caption{Distribution of the SI and TI for each source video clip in the MLVC-FRVQA dataset. The error bars represent the 95\% percentile range of SI and TI.}
  \label{fig:dataset:siti}
\end{figure}

We used the SI-TI Tool\footnote{\url{https://github.com/VQEG/siti-tools}} to calculate the Spatial perceptual Information (SI) and Temporal perceptual Information (TI) of all the source videos in the MLVC-FRVQA dataset. The SI indicates the amount of spatial detail in a frame, and TI refers to the amount of temporal changes in a video sequence \cite{itu-t_recommendation_p910_subjective_2021}. Figure~\ref{fig:dataset:siti} illustrates the average SI and TI distribution and 95\% percentile range per video. Besides the MS-Webcam set, the average SI and TI for all videos cover a wide range of spatial and temporal information. The new MS-Webcam dataset focuses on head and shoulder scenarios, mainly representing the video conferencing use case, showing a lower range of TI values.

\section{Method}
\label{sec:method}

\begin{figure}
    \centering
  \includegraphics[width = 0.6\columnwidth]{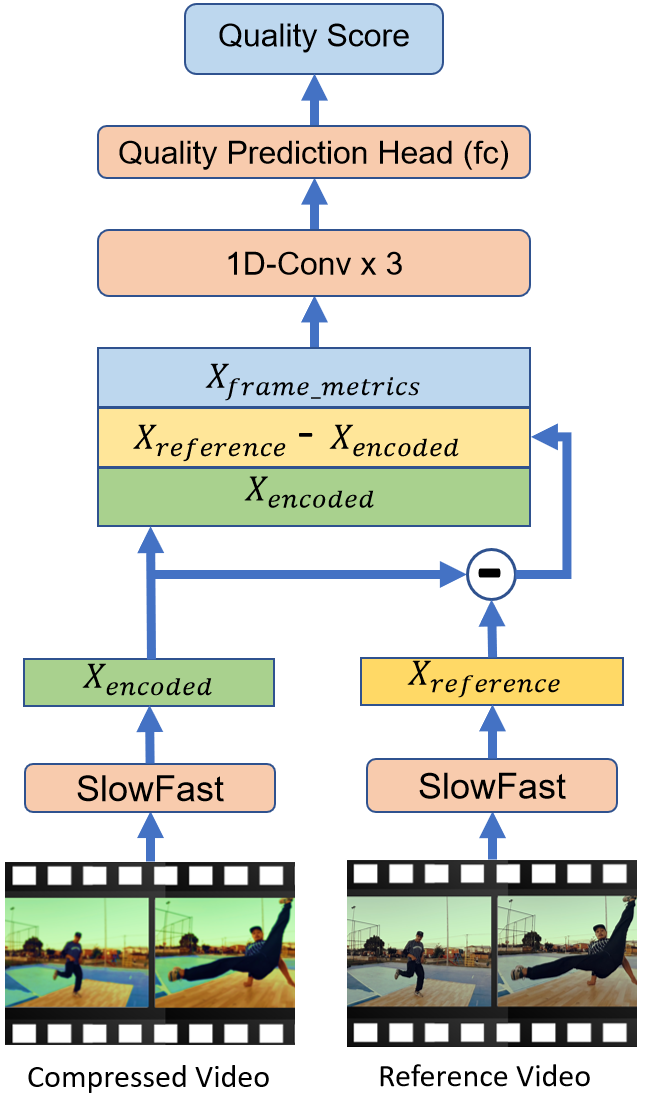}
  \caption{MLCVQA model architecture. The inputs for the model are SlowFast features ($T\times4608$) and frame-level metrics ($T\times352$), concatenated along the channel axis.}
  \label{fig:model_arch}
\end{figure}
We propose the Machine Learning Codec Video Quality Assessment (MLCVQA) FRVQA model, which involves a two-step approach: a feature extraction step, followed by a temporal quality prediction using the extracted features. This extends the ideas proposed in \cite{zhangUnreasonableEffectivenessDeep2018} and \cite{tariqWhyAreDeep2020} which show the effectiveness of using deep features as a perceptual metric for images, to the domain of videos. Note the feature extractor is frozen during the entire process (train and test), and the weights are never updated. Different from previous learning-based approaches which specifically mention avoiding augmentations other than random crops\cite{xu_c3dvqa_2020} while aggressively resizing the input videos, we propose relative perceptual quality invariant (RPQI) augmentations that do not affect the perceptual quality of a compressed video relative to its original version. We show that our simple approach with RPQI augmentations outperforms the current baselines. 

\subsection{Feature Extraction}
We choose the SlowFast \cite{feichtenhoferSlowFastNetworksVideo2019} model pre-trained using the action recognition task on the Kinetics-400 \cite{carreiraQuoVadisAction2017} for feature extraction. This choice was motivated by: (1) the fully convolutional nature of the model, allowing input videos of arbitrary resolution, and (2) the design of SlowFast being inspired by the Human Visual system.

We extract the features using the SlowFast model pre-trained on Kinetics \cite{carreiraQuoVadisAction2017}. Specifically, we use the SlowFast 8x8 with a ResNet-101 \cite{heDeepResidualLearning2016} backbone available on PyTorch Hub.

For each video, we fed a sliding window of 32 frames (sampled with a step size of 2 from a chunk of 64 consecutive frames), with a stride of 16, extracting 2304-D features before the last fully connected layer. Each window thus covers $2.133$ seconds (64 frames / 30 fps)  of the video. 

As mentioned above, SlowFast is a fully convolutional model allowing us to forward each clip (sequence of frames), in full resolution ($1920\times1080$) without resizing or cropping. Avoiding resizing is critical because many distortions introduced by the ML video codecs are not noticeable if the videos are resized to a smaller resolution. This is especially true for the top-performing ML video codecs from the CLIC 2022 Challenge.

For each pair of encoded video ($X_{enc}$) and reference video ($X_{ref}$) features, we calculate the difference ($X_{diff}$) as: 
\begin{equation}
    X_{diff}^{T\times2304} = X_{enc}^{T\times2304} - X_{ref}^{T\times2304}
\end{equation}
We further concatenate this difference to the encoded video features resulting in the concatenated feature vector ($X_{SF}$):
\begin{equation}
    X _{SF}^{T\times4608} = X_{enc}^{T\times2304} \mathbin\Vert X_{diff}^{T\times2304}
\end{equation}
Theoretically, the model should learn to model the difference on its own, but, in our experiments, we found providing the difference explicitly helps the model achieve better results as well as converge faster.

In our experiments, we noticed that in the few cases where a video does not have a lot of motion (i.e., it is more or less like a still image), then as one would expect, SlowFast features as well as learning-based baselines \cite{xu_c3dvqa_2020, sunDeepLearningBased2022} do not perform as well as frame-metric based methods such as VMAF (see Figure \ref{fig:TI-SI}). Thus, in order to make our model more robust to these outlier videos, we also input frame-level metrics. We pool frame-level metrics such as Visual Information Fidelity (VIF) \cite{vuViS3AlgorithmVideo2014} and Detail Loss Metric (DLM) \cite{liImageQualityAssessment2011}, in a similar manner as the sliding window approach described above. For each window of 32 frames, we end up with $32\times11$ frame level metrics which are flattened to a vector of size 352. This vector is concatenated with the SlowFast features resulting in the final feature representation ($X$) of the video pair which is fed to the model:
\begin{equation}
    X^{T\times4960} = X_{SF}^{T\times4608} \mathbin\Vert X_{frame\_metrics}^{T\times352}
\end{equation}

\subsection{Data Augmentation}
In order to increase the data size, we propose augmentation strategies that do not affect a clip's perceptual quality relative to a reference clip. For spatial augmentation, we extract features using different transforms such as horizontal and vertical flips, and rotation by small angles 
($|\theta| \in  [0^{\circ},5^{\circ},10^{\circ}]$). We also add a center crop ($500\times500$) augmentation to implicitly give more weight to the center region of each frame, since the human visual system often pays more attention to the center of its view\cite{zhangStudySaliencyObjective2017}, thus having more impact on the perceived quality of the video.

Before settling on the final set of augmentations, we explored different strategies, including modifying the distorted videos at the just noticeable difference (JND) level, and 3x the JND level. For these experiments we included changes in brightness, contrast, hue, saturation, gamma, Gaussian noise, rotation, and horizontal flipping, parameter levels were defined through crowd-sourced P.910 DCR subjective tests. These experiments are discussed in Section \ref{sec:augmentation} in more detail and did not improve the model's performance. 

We also augment (temporal sample) the data along the temporal axis. Specifically, given a $T\times2304$ feature representation of a video, we sample feature vectors with a step size of 2, resulting in two features of size $[T/2]\times2304$.

We follow the standard practices of training augmentation \cite{xu_c3dvqa_2020, feichtenhoferSlowFastNetworksVideo2019, carreiraQuoVadisAction2017, shanmugamBetterAggregationTestTime2021}. During training, random augmentations are applied to both the encoded video as well as the reference video. The resulting features are then sampled temporally as mentioned above. It should be noted that the same augmentation is applied to both the encoded and reference video to ensure the relative perceptual quality stays the same. We also experimented with test-time augmentation, but it showed no improvement in performance.

\subsection{Model Architecture}

An overview of the model is shown in Fig.~\ref{fig:model_arch}. Our model consists of a projection layer followed by two 1-D convolutional layers and a quality prediction head.

The projection layer consists of a 1-D convolutional layer with a
kernel size of 3, a stride of size 1, and padding of size 1. This layer projects the input 4608-D features to a size of 128-D. The two subsequent convolutional layers have the same configuration with the only difference being that of keeping the number of input and output channels the same, viz. 256-D. ReLU activation functions follow each convolutional layer.

We chose the kernel size of $3$ to help incorporate information from neighboring features. Setting stride and padding to $1$ ensures the temporal dimension does not change during convolutions.

After the 1-D convolutions is a 2-layer multi-layer perceptron to predict the score at each time step. This allows a fine-grained quality score prediction since the quality of a video clip can be dynamic over time. Finally, we average (arithmetic mean) the scores across the time dimension to get the final predicted score for the video. 

We use Smooth L1 loss to model the error between the predicted quality score ($\hat{y}$) and the ground truth MOS ($y$):

\begin{equation}
L(y,\hat{y})=
    \left\{\begin{matrix}
        \frac{1}{2}(y - \hat{y})^{2} & if \left | y - \hat{y}  \right | < 1\\
        \left | y - \hat{y}\right | - 0.5 & otherwise
    \end{matrix}\right.
\end{equation}

\subsection{Training details}
We optimize the model using the Adam optimizer with a batch size of $32$, weight decay set to $10^{-4}$, and momentum of $0.9$. The maximum learning rate
is set to $4 \times 10^{-4}$ and a linear warm-up is applied in the first $10$
epochs. After $10$ epochs the learning rate is decayed using the cosine decay schedule. We train for $200$ epochs after warm-up, which takes about $75$ minutes on 8 NVidia V100 GPUs \footnote{We pre-extract the features to speed up the training process. Without pre-extraction, we expect the training to take longer.}.

\subsection{Evaluation metrics}
\label{sec:metrics}
\begin{figure}
\centering
  \includegraphics[width= 0.7\columnwidth]{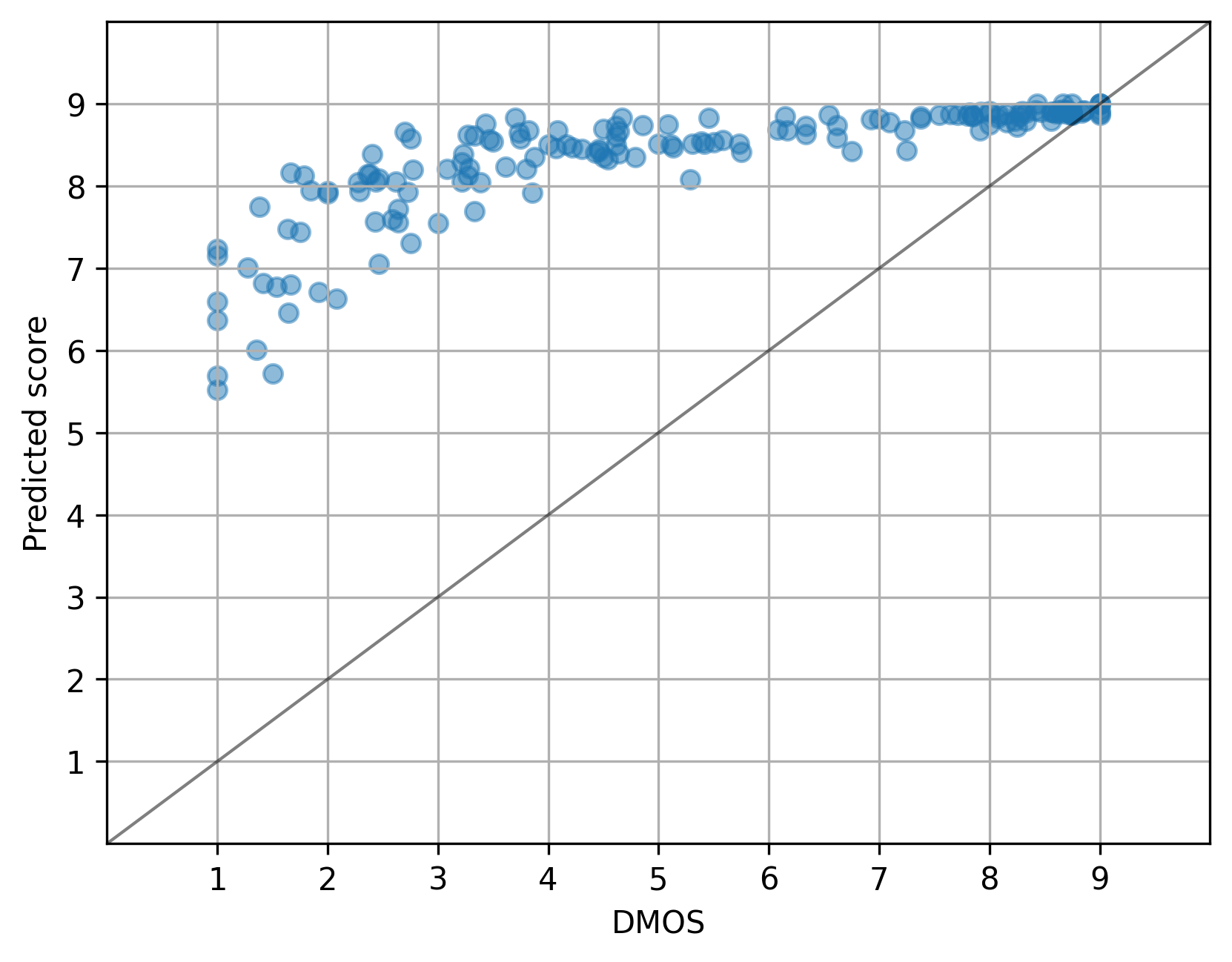}
  \caption{This plot shows the superiority of Tau-b over SRCC as a metric for comparing the quality rankings of models. The figure is an example of an FRVQA model with an inaccurate prediction of the subjective score, but a high SRCC (SRCC = 0.92 and Tau-b = 0.75).}
  \label{fig:srcc}
\end{figure}

A key performance indicator for an FRVQA model is the validity of the rank order of ML video codecs produced by it and compared to the subjective test rank order. Typically, SRCC and Kendall's Tau ($\tau$) \cite{kendall_new_1938} are used for measuring the ordinal association between two sets of scores (i.e., DMOS values from a subjective test and predicted scores from an FRVQA model). The SRCC represents the linear relationship between two rankings, for which a high coefficient can be achieved when the two ranked sets are monotonically related. Kendall's Tau measures how much two ranking procedures agree for all combinations of item pairs in the set. 
It is well known that Kendall's Tau is a better estimate of correlation in population and is the preferred method when there is a small number of items, and many tied ranks \cite{howellStatisticalMethodsPsychology,fieldDiscoveringStatisticsUsing2022}. It also provides a better interpretation compared to the SRCC.
$\tau = t$ shows that for a randomly selected pair of items (e.g., ML video codecs), the probability that the two items will be ranked in the same order as the subjective test is $t$ higher than the probability that they will be ranked in the reverse order. In other words, it shows how far the conclusions made from the prediction of an objective model correspond to the conclusions made from the subjective test.
Fig.~\ref{fig:srcc} illustrates the prediction of a poor-performing FRVQA model compared to the subjective scores resulting in SRCC $= 0.92$ and $\tau= 0.75$. Consequently, a high SRCC score, in this case, can be a misleading metric.

Furthermore, none of these methods consider the uncertainty in subjective scores. Both methods consider two items to construct a tied rank only when they have equal numeric values. However, the  MOS from a subjective test represents the sample mean and is subject to uncertainty, i.e., the numeric MOS value can change by a minor change in the rating of a single participant. Therefore, it is recommended to consider the distribution of ratings (e.g., through 95\% CI ranges) when comparing MOS values of two items \cite{naderiTransformationMeanOpinion2020, seufertFundamentalAdvantagesConsidering2019}. 
For our evaluation, first, we create a rank order of the items given the DMOS and the 95\% CI values from the subjective test according to \cite{naderiTransformationMeanOpinion2020}, i.e., two items considered a tied rank when the DMOS value of one item is in the 95\% CI range of the other item. Afterward, we calculate Kendall's Tau-b on the provided ranked-order list (hereafter referred to as Tau-b 95).
Given the higher number of tied ranks, a small number of ML video codecs, and better interpretability of Kendall's Tau, we consider Tau-b 95 the primary metric in our evaluation.

Besides ranked-based metrics, we also calculate Pearson Correlation Coefficient (PCC) and Root Mean Square Error (RMSE) to evaluate the linearity and accuracy of predictions, respectively.
\subsection{Number of votes and expected accuracy}
\label{bootstrapping}

\begin{figure}[tb]
    \centering
    \subfloat[\label{fig:vbote:coef:model}]{\includegraphics[width=0.25\textwidth]{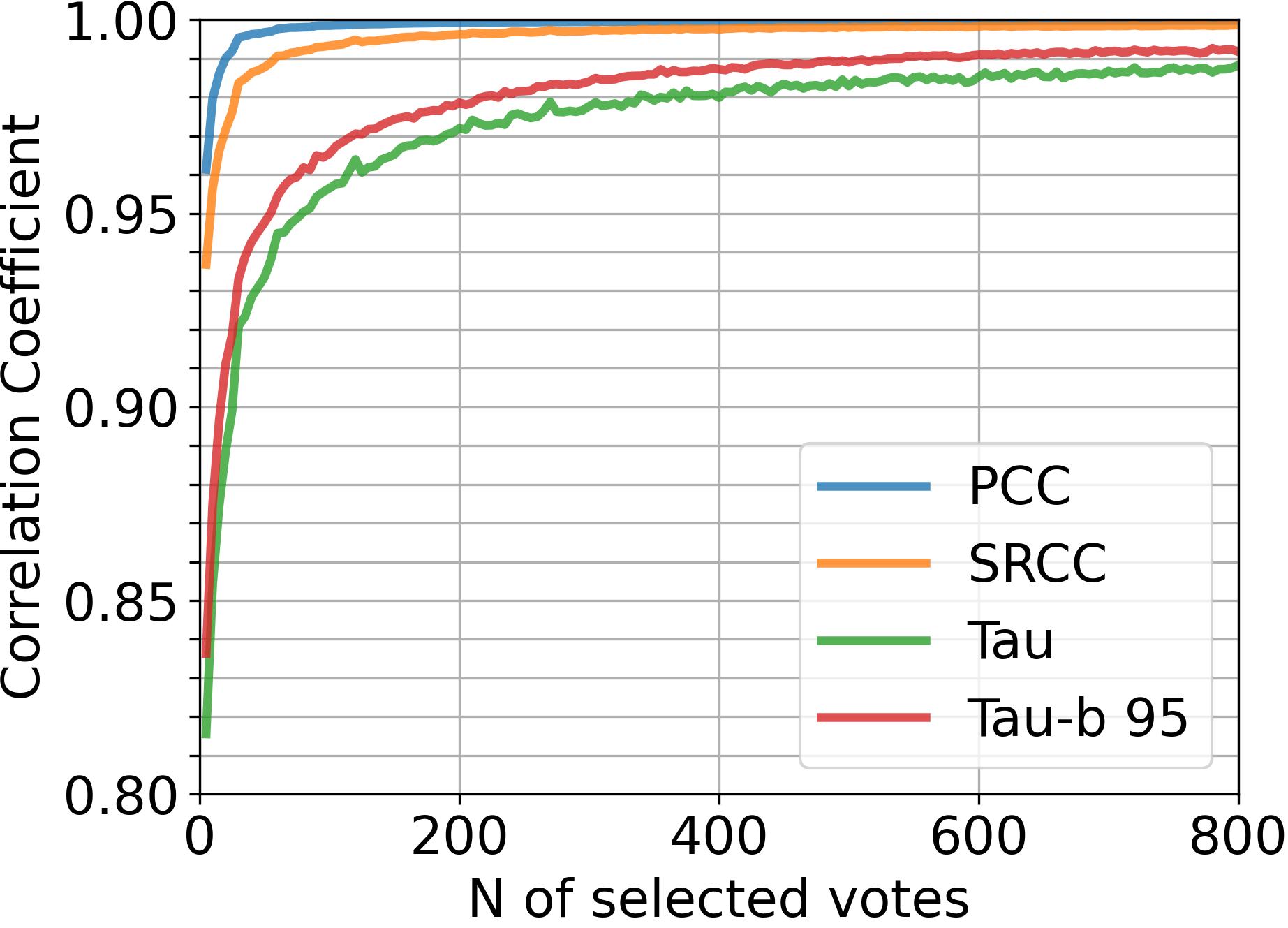} }
    \subfloat[\label{fig:vbote:coef:clip}]{\includegraphics[width=0.25\textwidth]{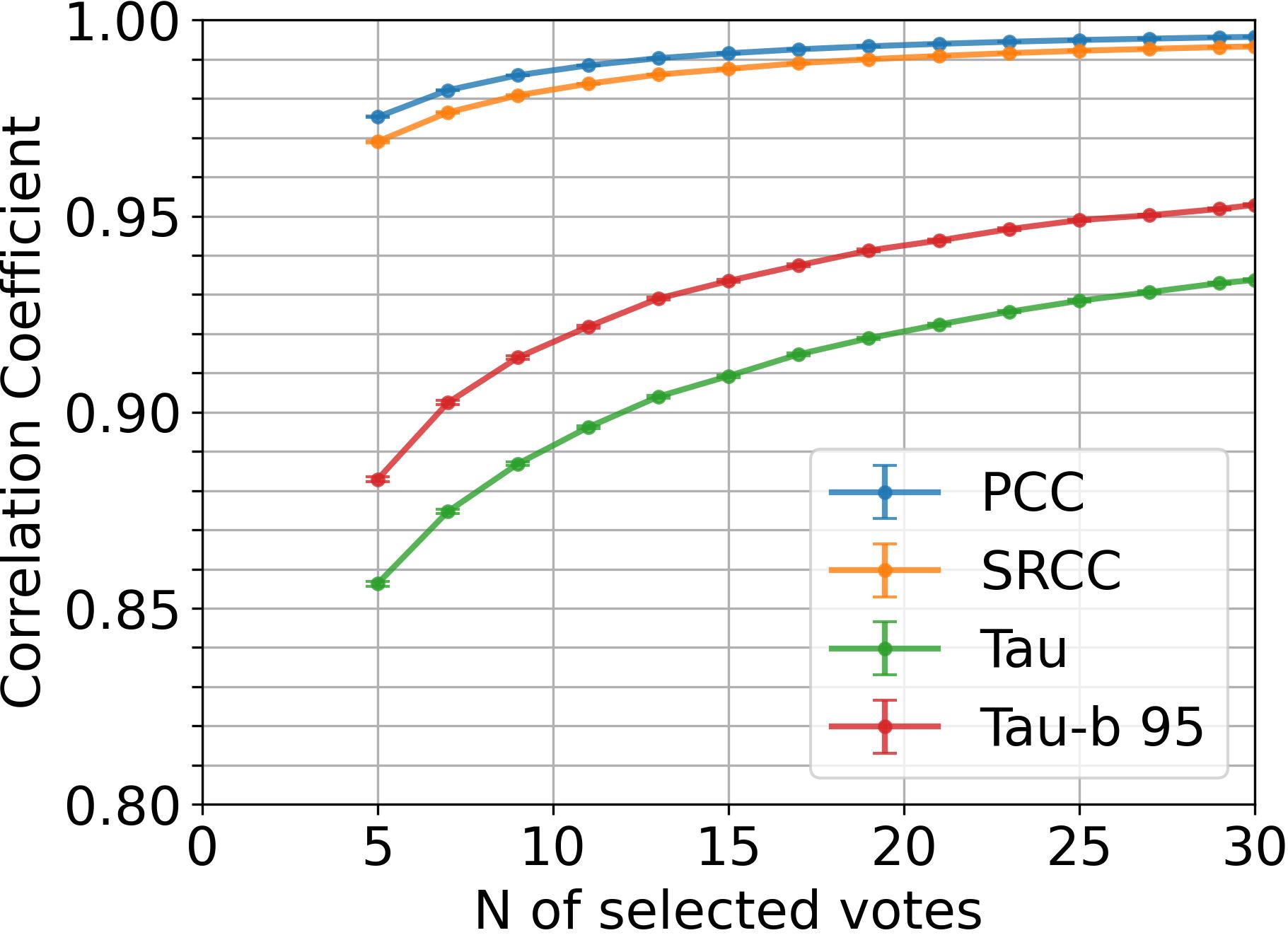} }

    \caption{Changes of different statistics by increasing the number of votes used for calculating DMOS. (a) clip level, and (b) model level. Error bars represent 95\% CI during 200 simulation runs. }

    \label{fig:n_votes}
\end{figure}    

\begin{figure*}[tb]
    \centering
   
    \subfloat[]{\includegraphics[width=0.275\textwidth]{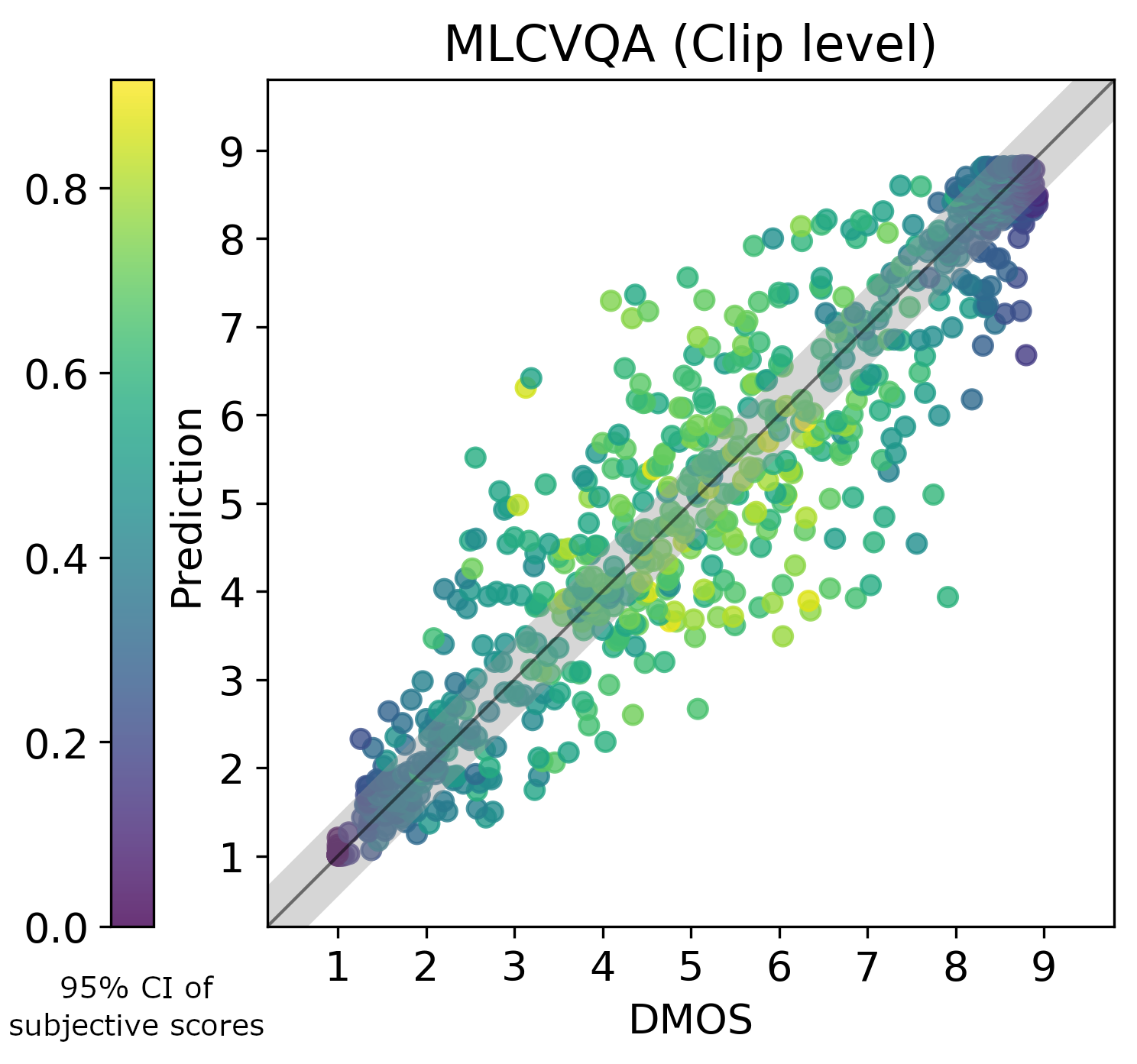}}
    \subfloat[]{\includegraphics[width=0.24\textwidth]{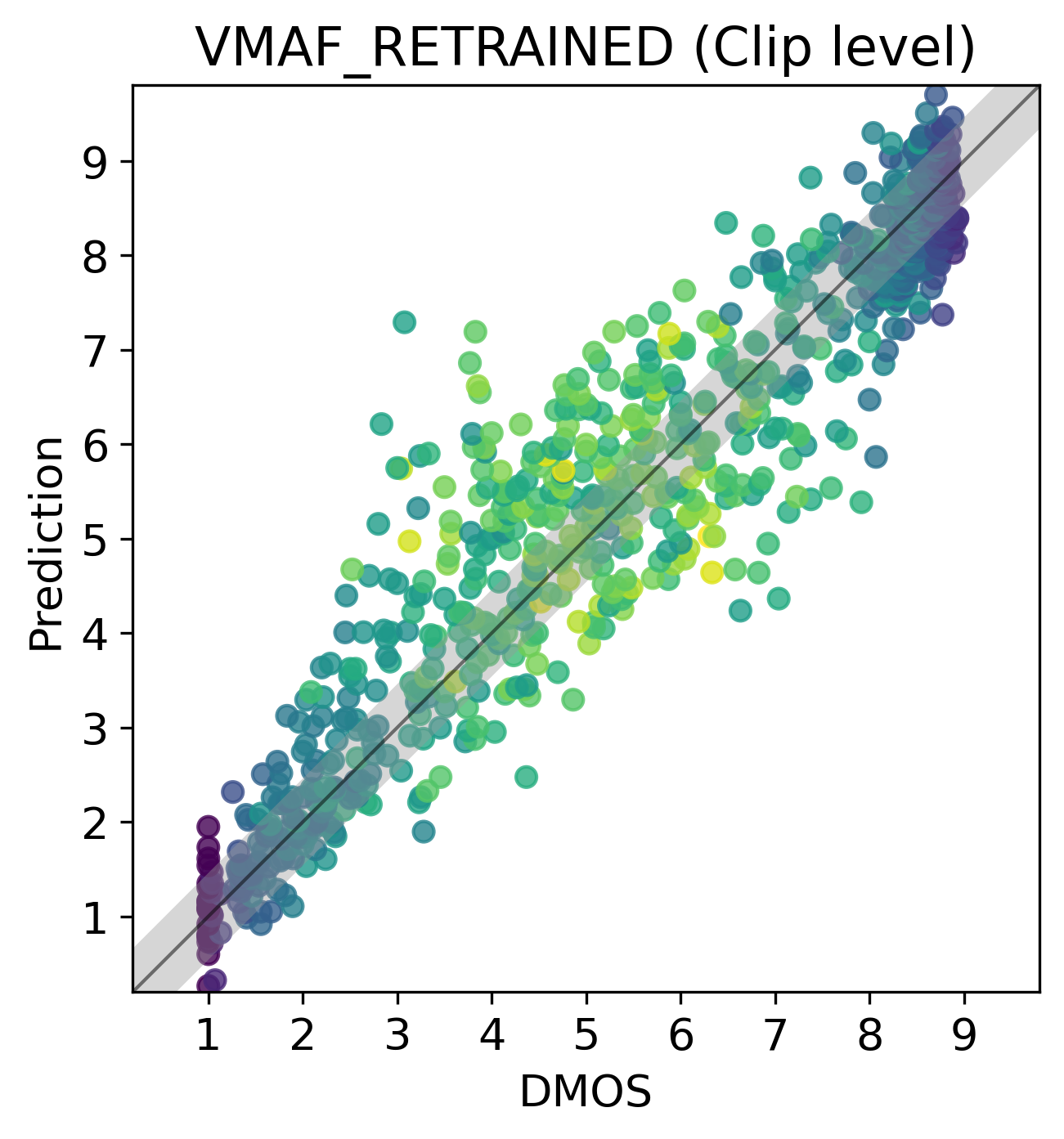}}
    \subfloat[]{\includegraphics[width=0.24\textwidth]{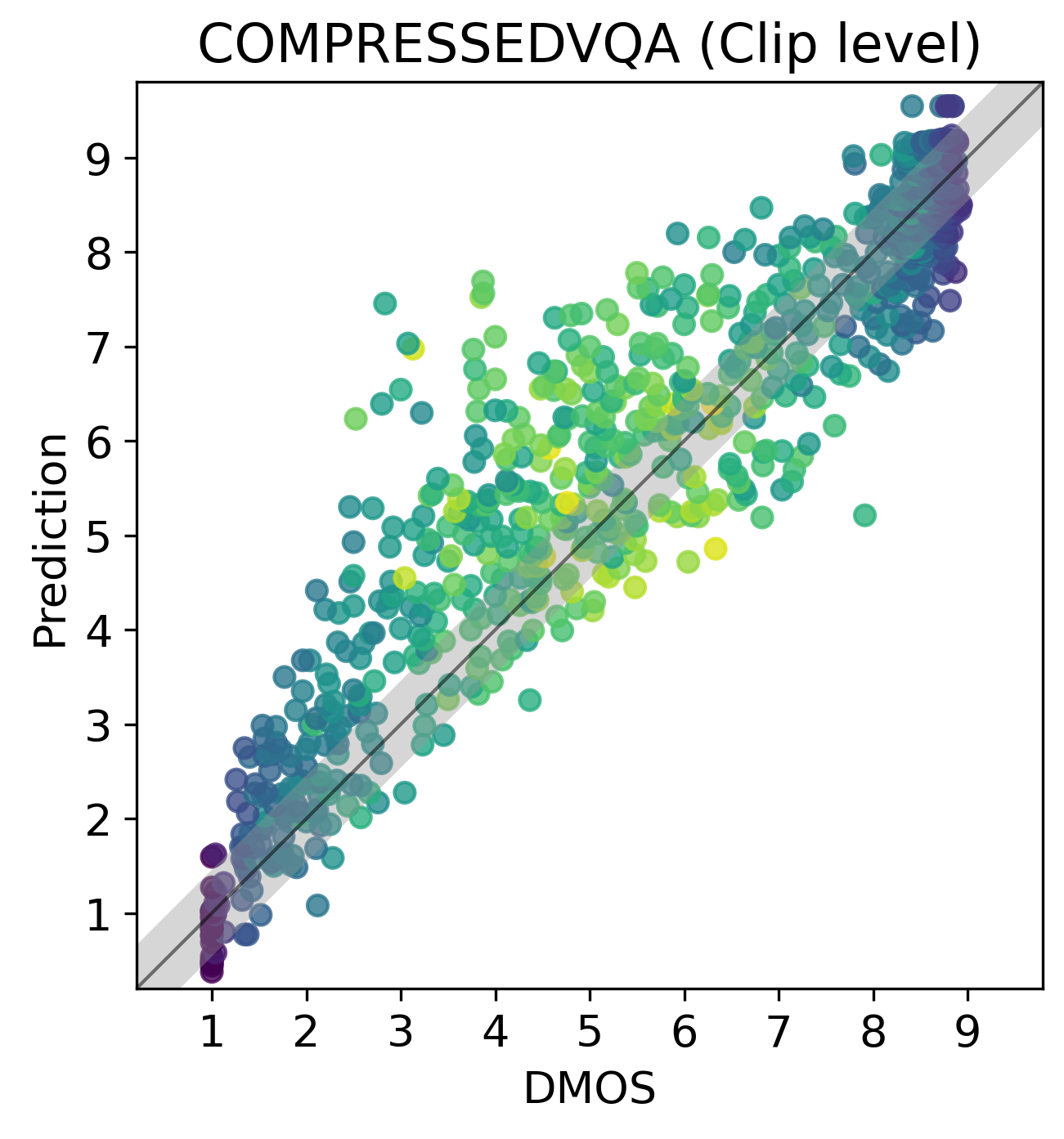}}    
    \subfloat[]{\includegraphics[width=0.24\textwidth]{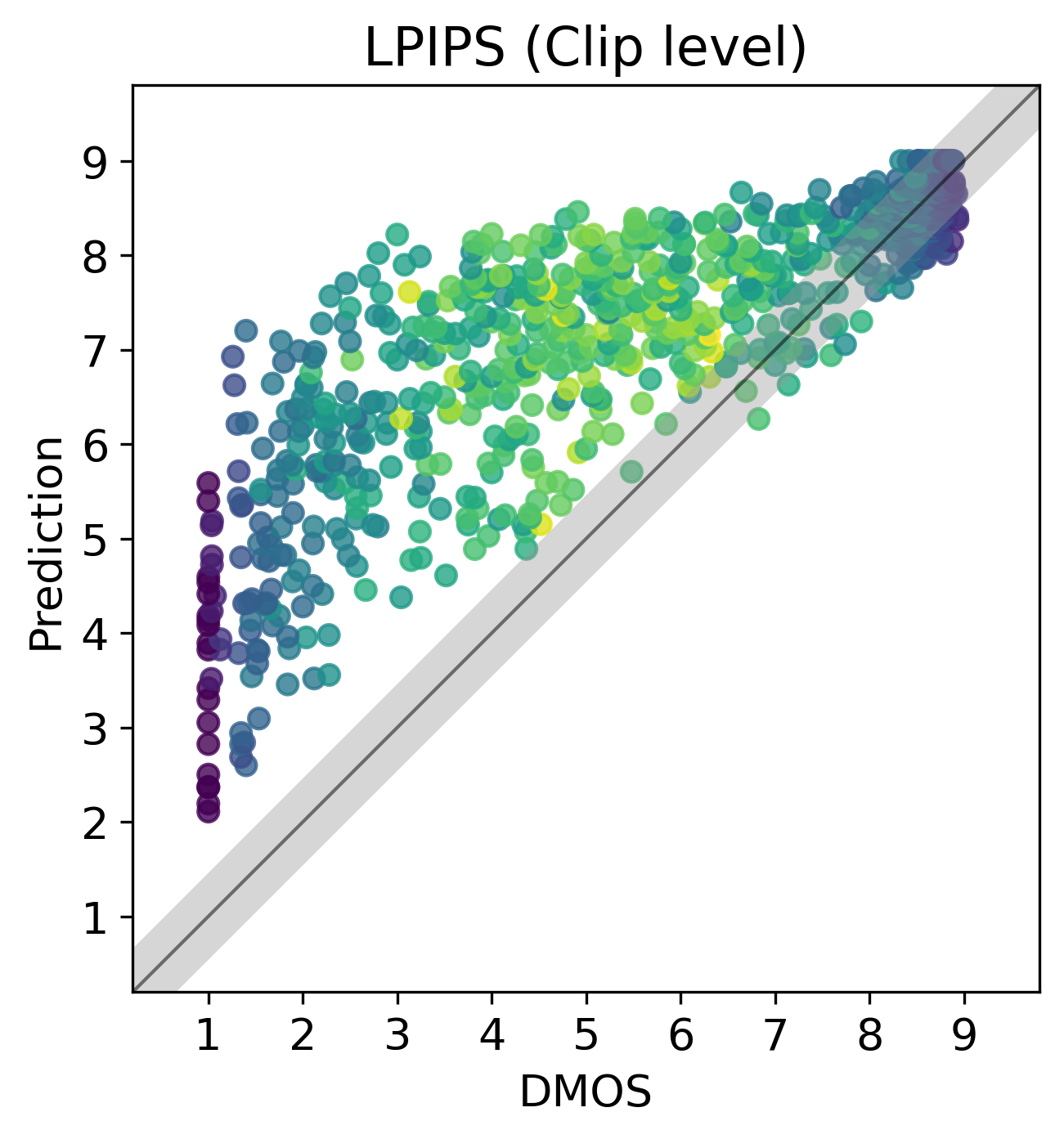}}  
    \\
    \subfloat[]{\includegraphics[width=0.24\textwidth]{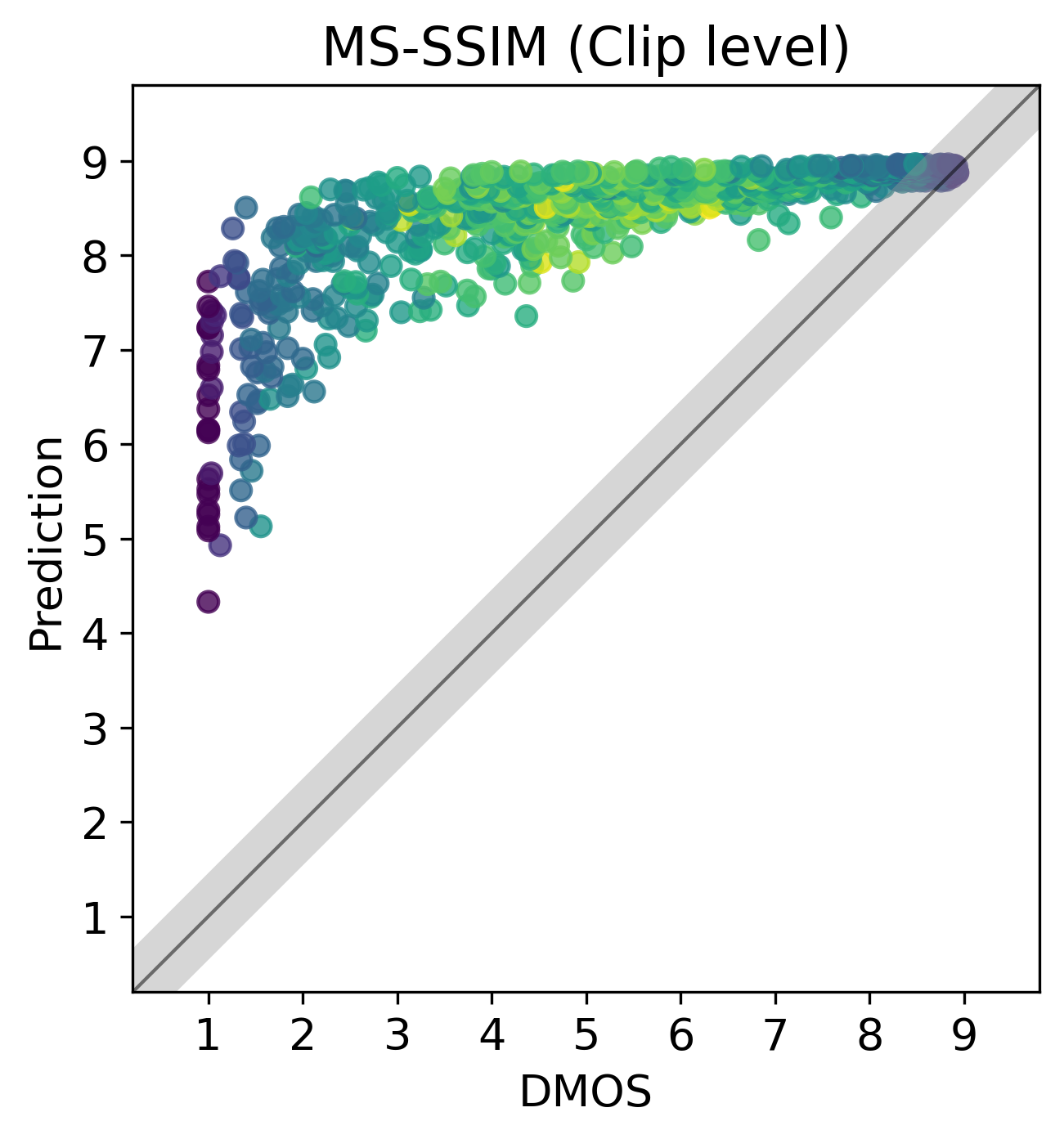}}
    \subfloat[]{\includegraphics[width=0.24\textwidth]{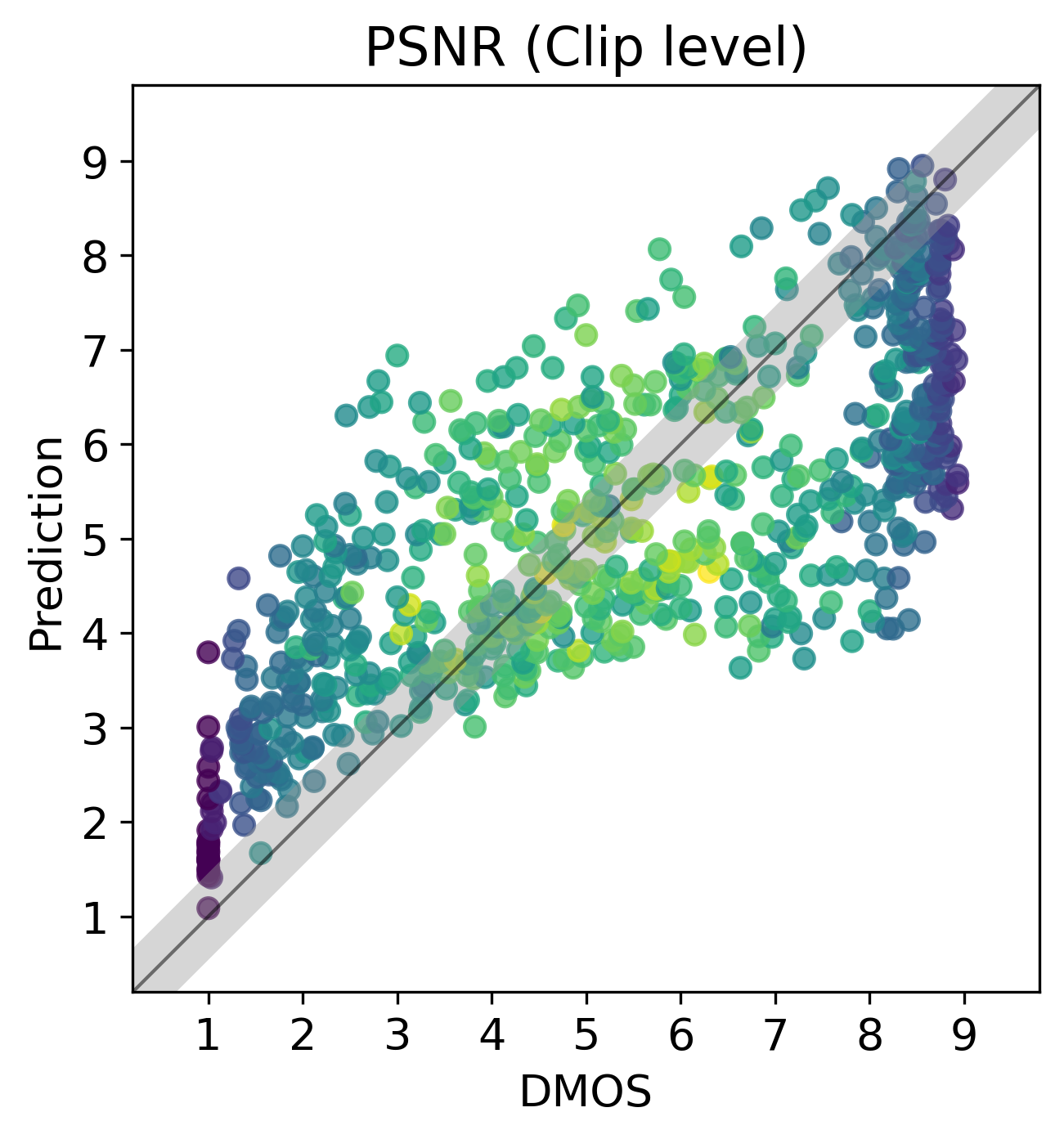}}
    \subfloat[]{\includegraphics[width=0.24\textwidth]{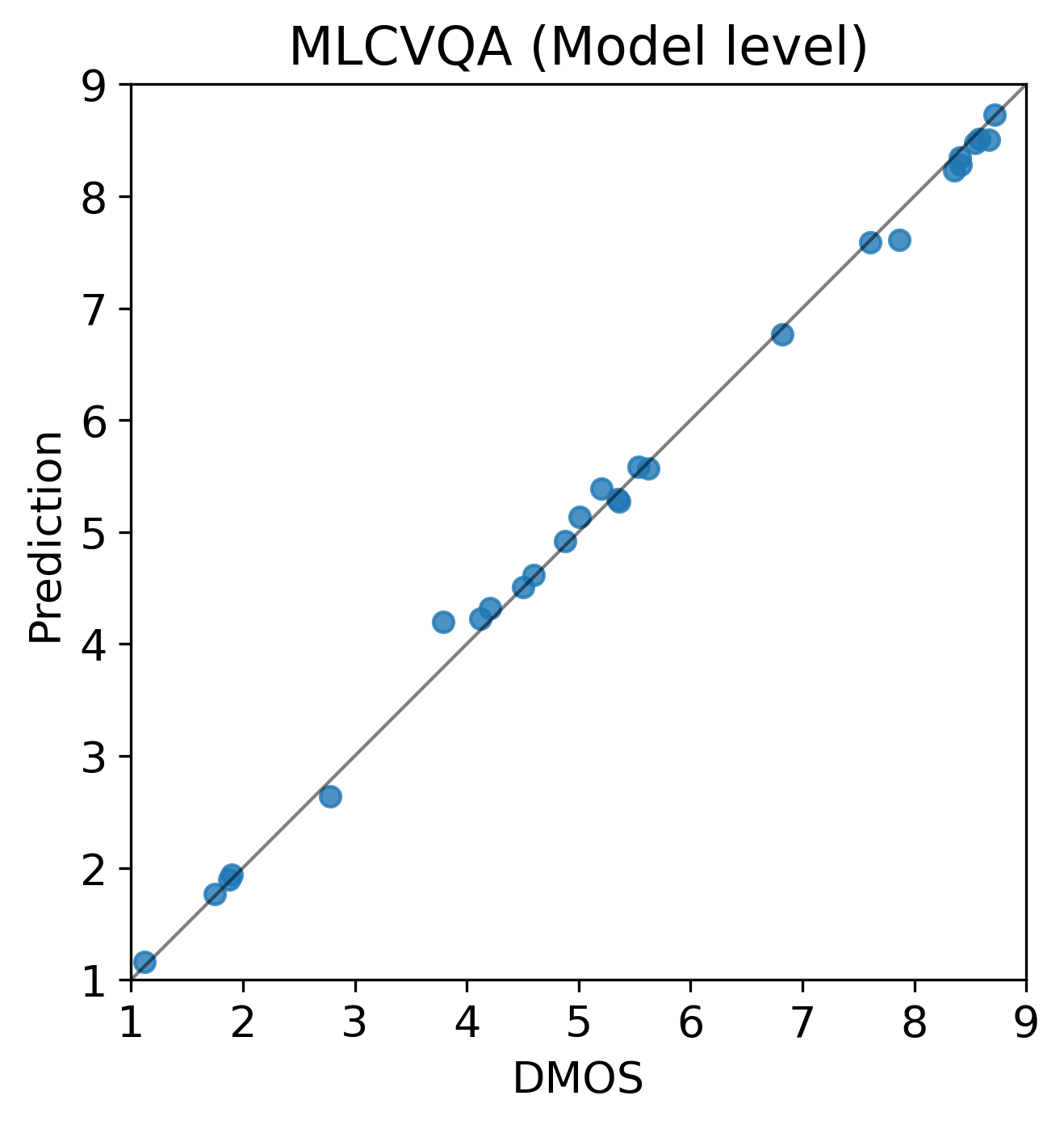}}
    \subfloat[]{\includegraphics[width=0.24\textwidth]{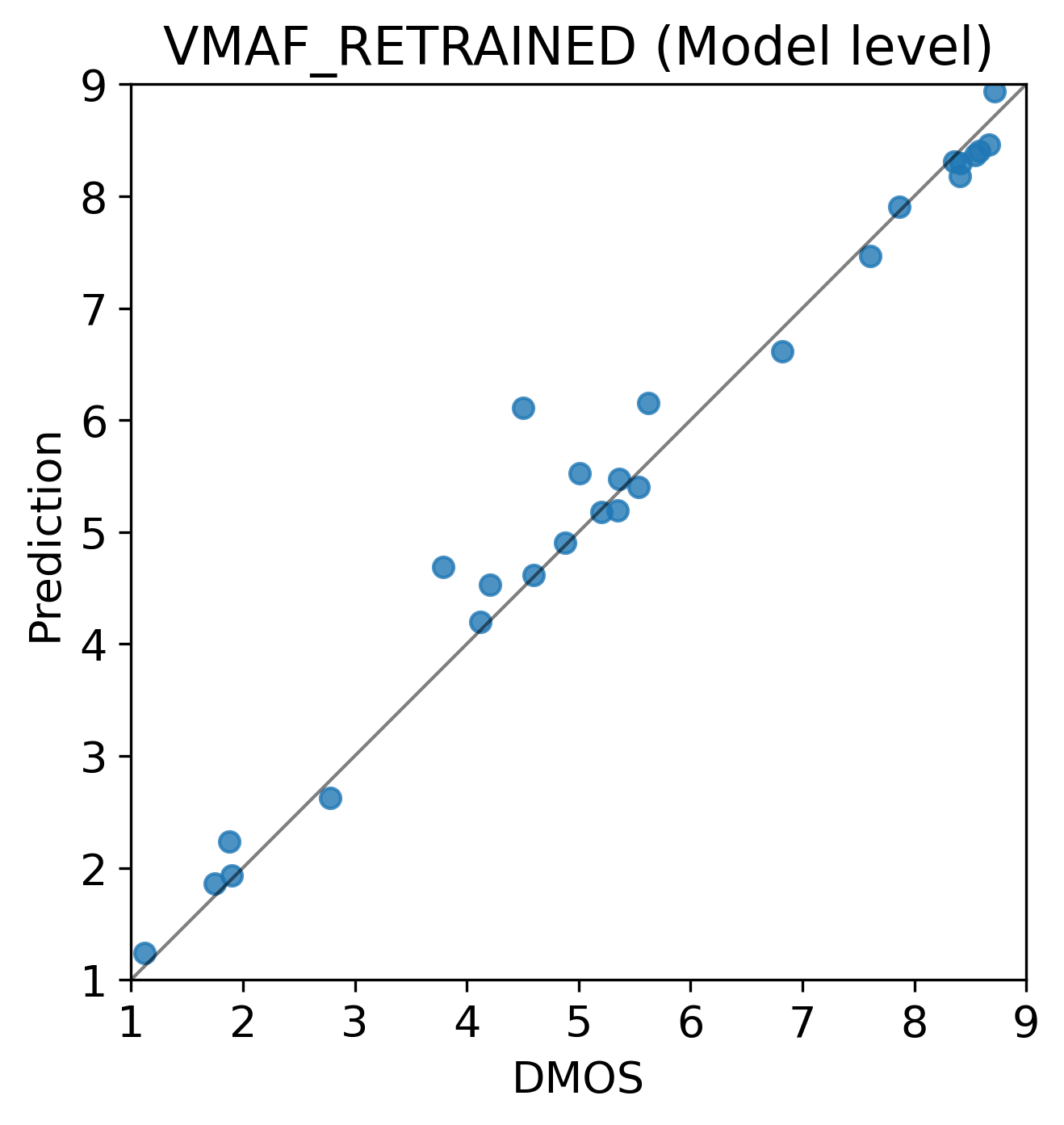}}
    \\
    \subfloat[]{\includegraphics[width=0.27\textwidth]{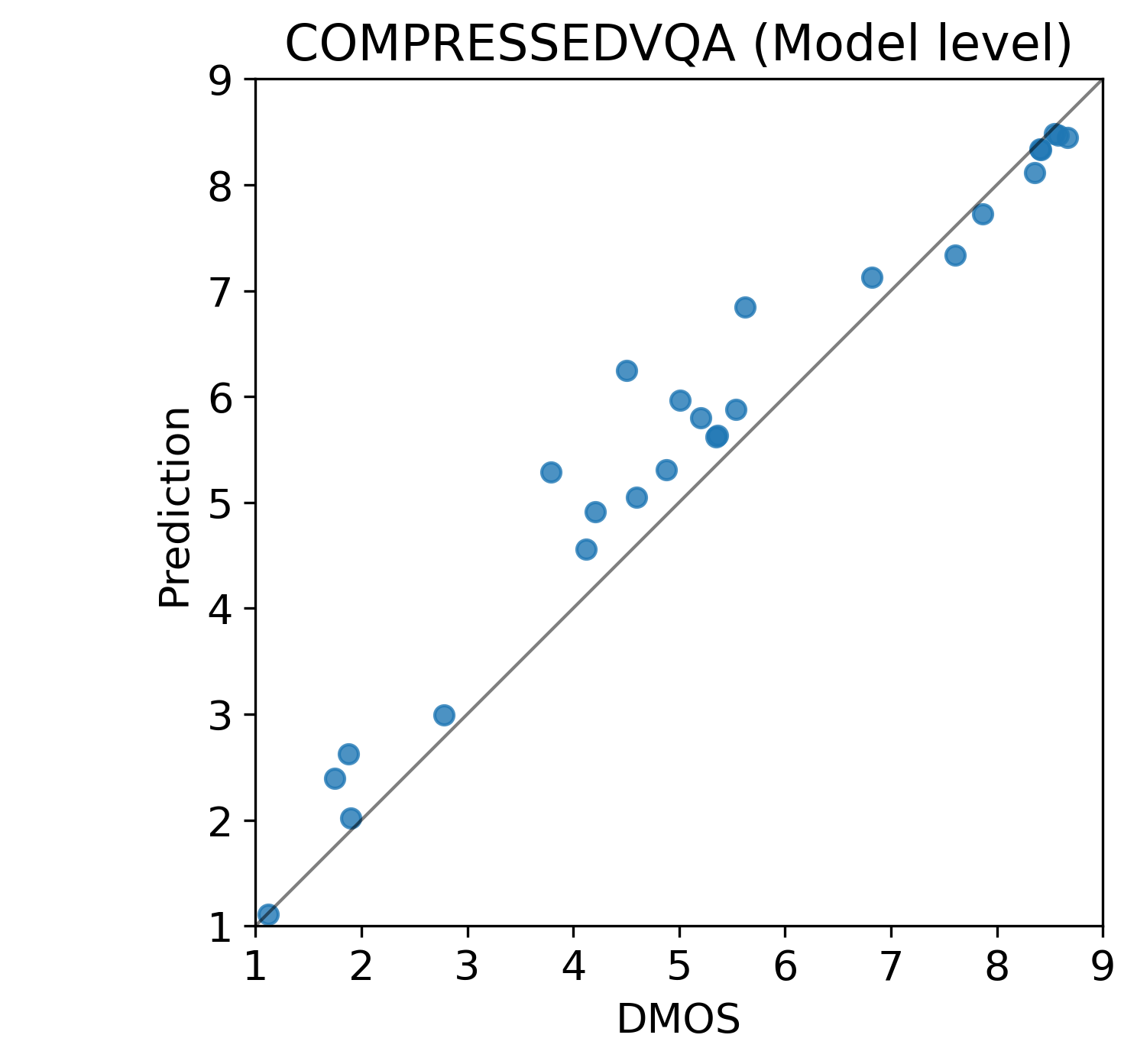}}    
    \subfloat[]{\includegraphics[width=0.24\textwidth]{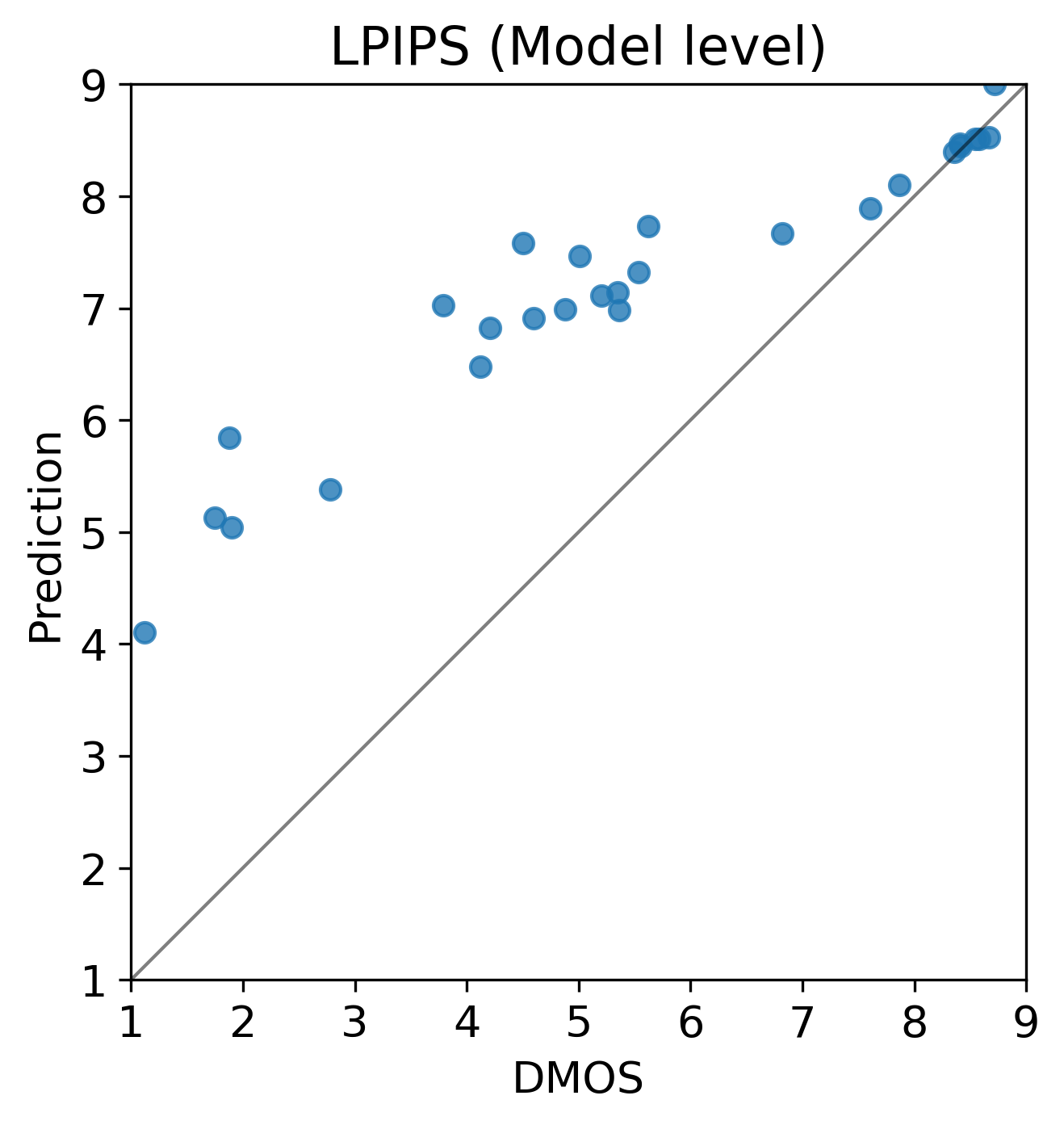}}
    \subfloat[]{\includegraphics[width=0.24\textwidth]{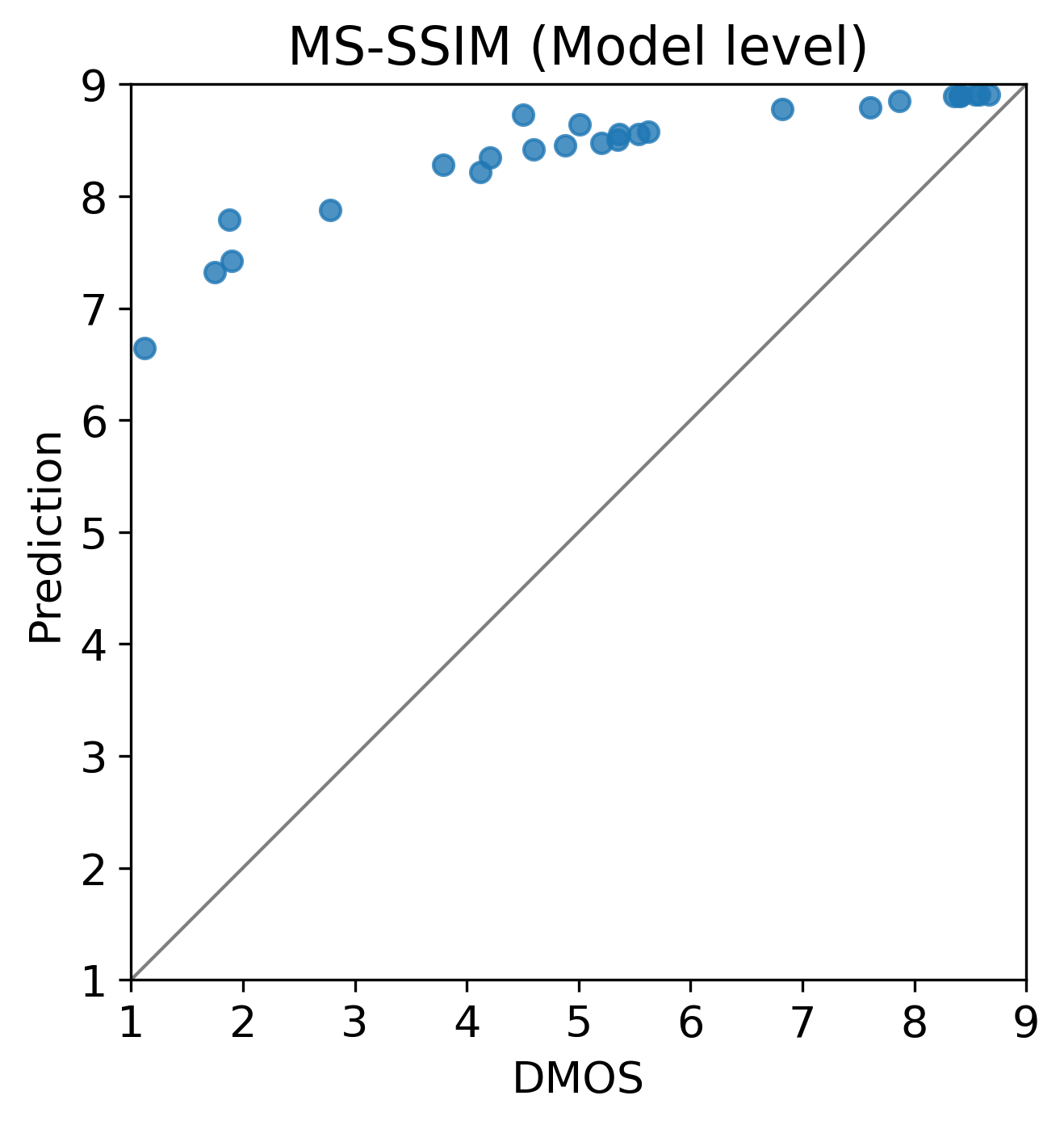}}
    \subfloat[]{\includegraphics[width=0.24\textwidth]{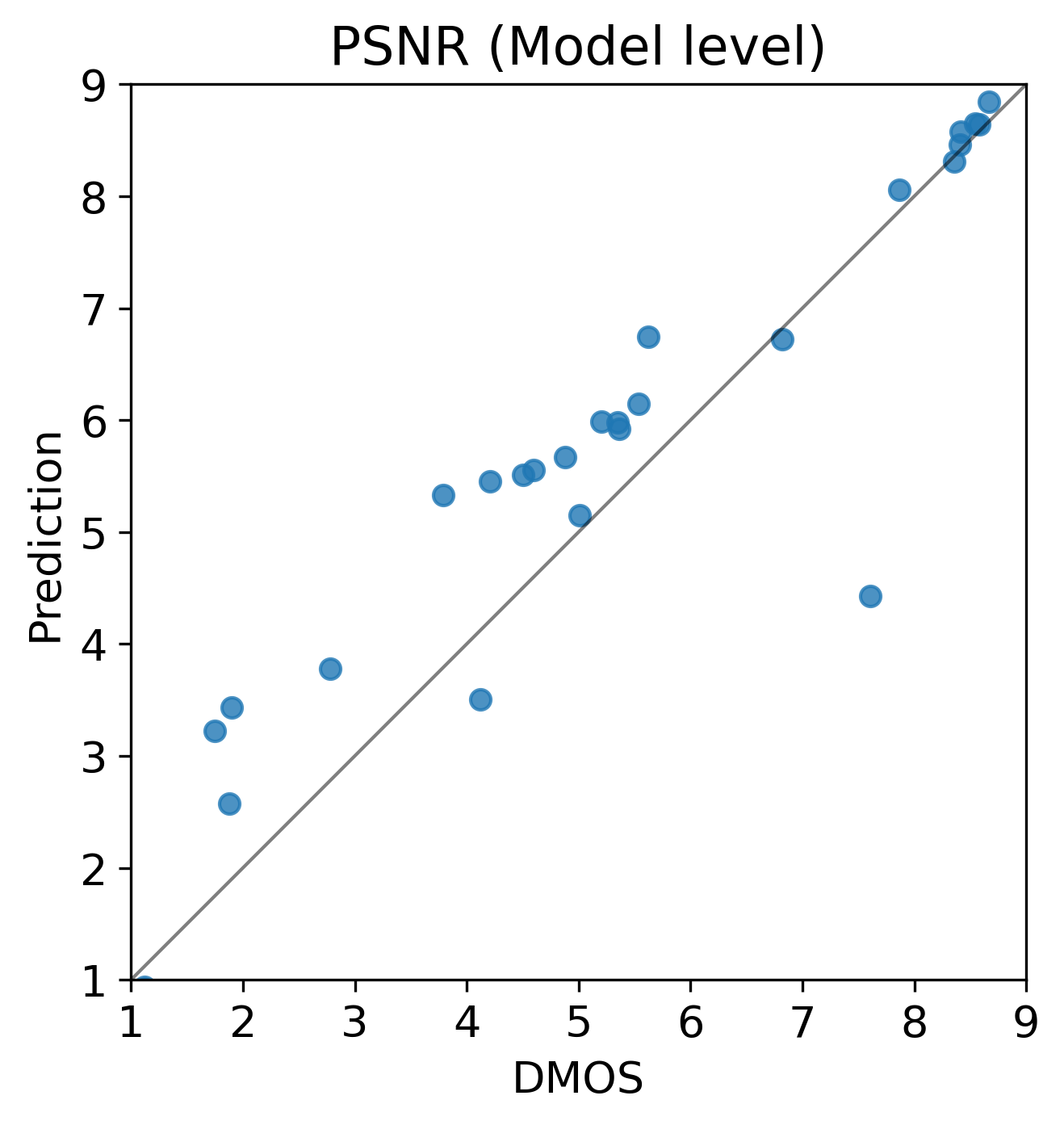}}
 
    \caption{Distribution of MLCVQA (a,g),  retrained VMAF (b, h), CompressVQA (c, i), LPIPS (d, j), MS-SSIM (e, k), and PSNR (f, l) predictions in clip and model levels, respectively. Predictions of LPIPS, MS-SSIM, and PSNR are linearly scaled to the [1,9] range. In the clip level, lighter colors show larger 95\% CI width in the subjective test. Gray areas mark the average 95\% CI of subjective test around the diagonal line. }
    \label{fig:results:scatter}
\end{figure*}

We used a bootstrapping simulation to estimate the accuracy of DMOS scores as a function of the number of votes used per clip and model. In the simulation, N votes were randomly selected per clip (model) with replacement and used to calculate the DMOS values of that subset. Consequently, we calculated correlations and RMSE between DMOS from the subset and when the entire ratings are used. We repeated this process 200 times per the number of selected votes (N) and calculated the mean and 95\% CI for all metrics\footnote{The average 95\% CI for all metrics in model level was 0.001.}. Fig.~\ref{fig:n_votes} illustrates how each metric changes as a function of the number of votes used in the calculation of DMOS. At the model level, PCC is saturated by $\sim$100 votes, whereas the SRCC shows a slight increase up to $\sim$200 votes. Kendall’s Tau-b and Tau-b 95 partially saturated by 500 votes however with only a change of $\sim$0.01 compared to 200 votes. RMSE is also saturated by $\sim$600 votes although with only a 0.05 difference compared to 200 votes (see Figure \ref{fig:bstrp:rmse}). At the clip level, although PCC and SRCC are saturated by $\sim$25 votes, Tau-b has a steady slope within the range of simulation. Similarly, RMSE tends to decrease further when more votes are used. A similar comparison for ACR ratings showed that $\sim$40 votes per clip are recommended \cite{naderi_crowdsourcing_2022}.

Subjective scores have an uncertainty that decreases by increasing the number of votes collected in a dataset. One can conclude that the prediction accuracy of a hypothetical objective model, when compared to subjective scores, may only reach the limit observed during simulation, given the number of subjective votes collected for the test set. 
In this paper, we use five-fold cross-validation, in which 6 videos (encoded by 27 codecs) belong to the test set in each fold (details in section~\ref{sec:results}). Consequently, the predictions of the FRVQA models were compared to the subjective DMOS values calculated from $\sim$26 votes per clip and $\sim$160 votes per model.
Given the simulation result, the expected upper limit for each statistical metric is given in Table \ref{tab:max_expected_result}.

\begin{figure}[tb]
    \centering
    \subfloat[\label{fig:vbote:coef:model}]{\includegraphics[width=0.25\textwidth]{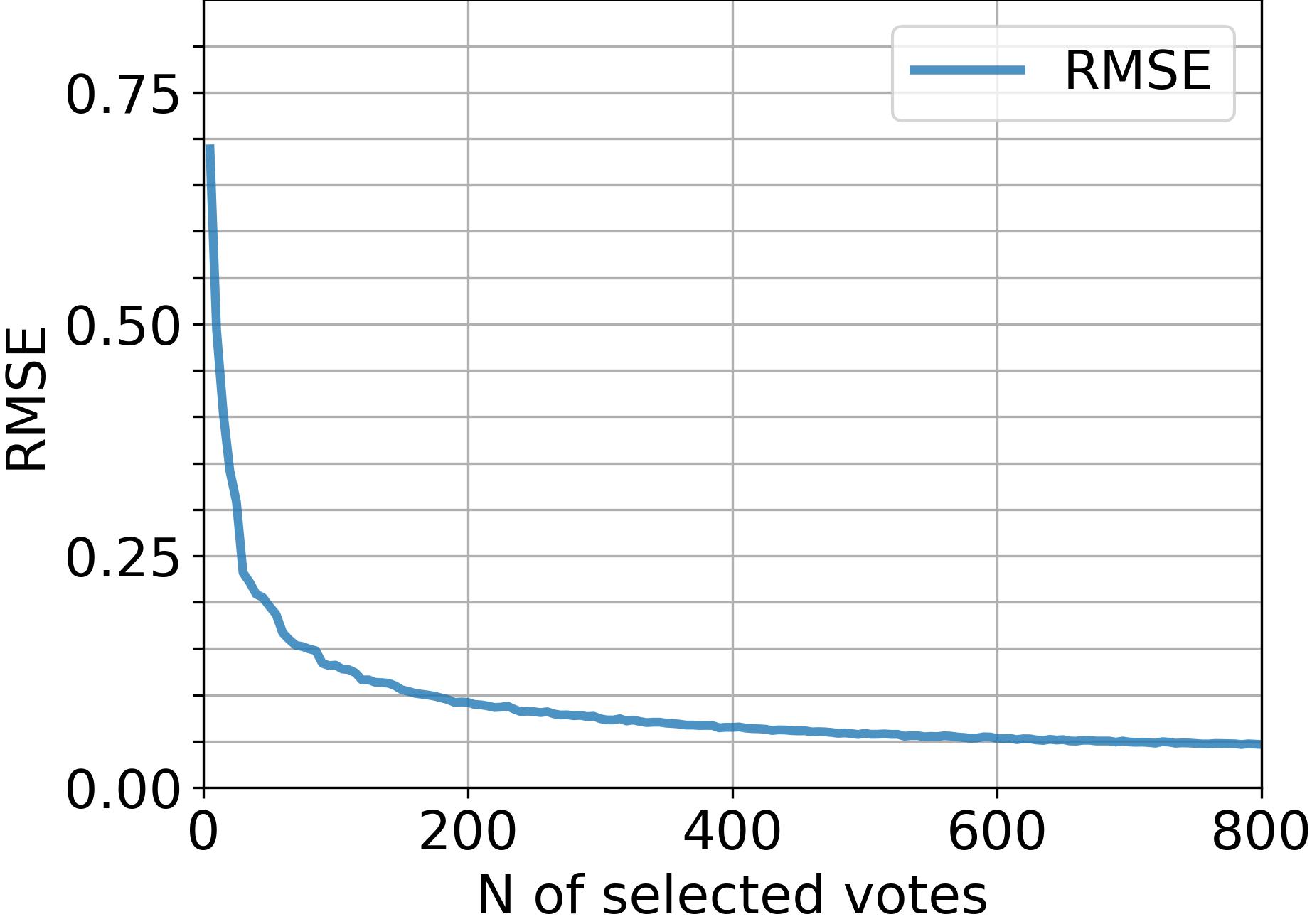} }
    \subfloat[\label{fig:vbote:coef:clip}]{\includegraphics[width=0.25\textwidth]{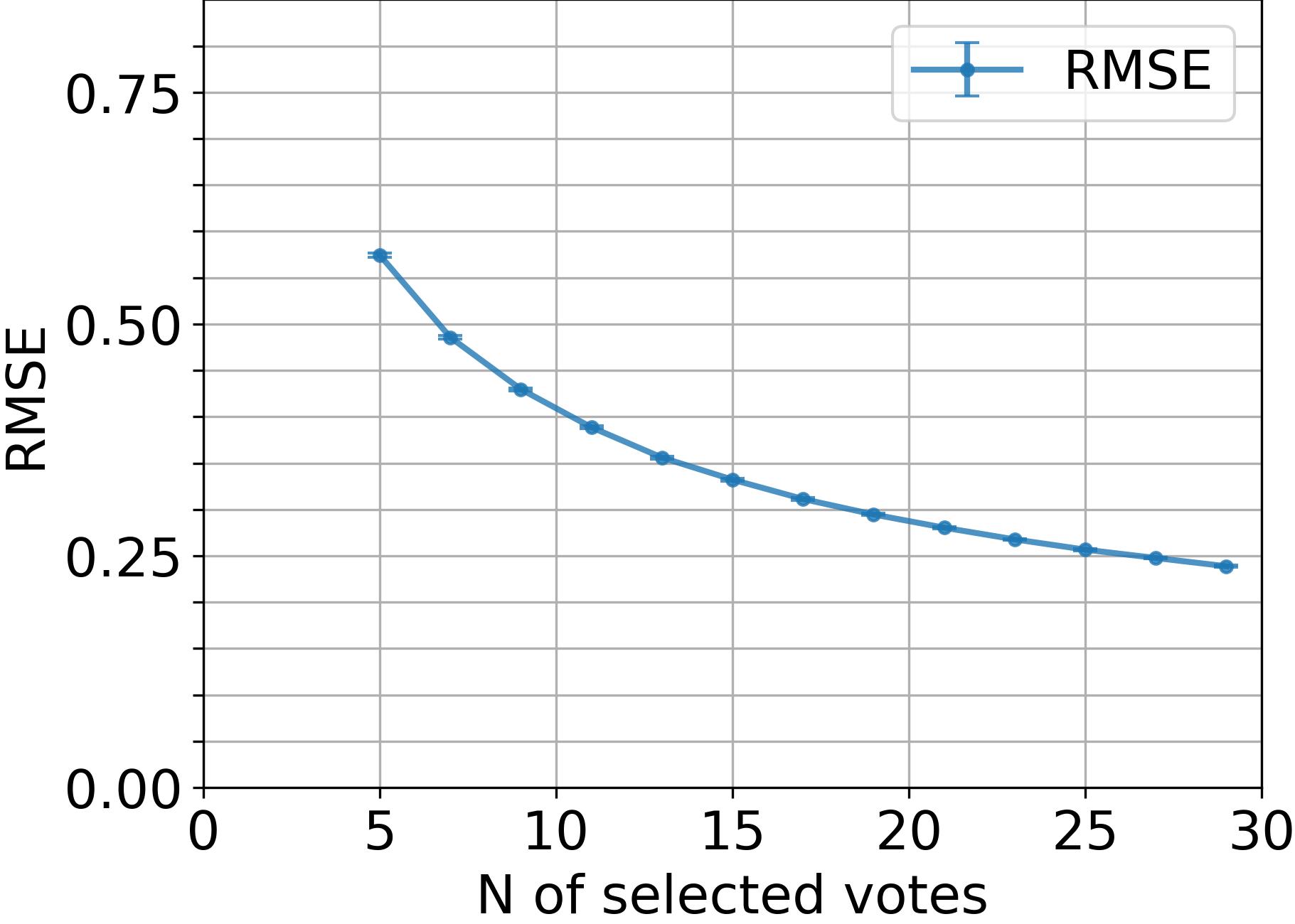} }

    \caption{Changes of RMSE by increasing the number of selected votes used for calculating DMOS. (a) clip level, and (b) model level.}
    \label{fig:n_votes_rmse}
\end{figure}

Figure~\ref{fig:n_votes_rmse} illustrates how the RMSE is changing as a function of the number of selected votes in the bootstrapping simulation. 

\begin{table}[htbp]
    \caption{The upper limit accuracy of a hypothetical objective model inferred from bootstrapping given the number of subjective ratings used in the test sets (26 votes per clip and 160 per model level).}
    
    \label{tab:max_expected_result} 
    \begin{center}
    \resizebox{0.8\columnwidth}{!}{%
        \begin{tabular}{ c  c c c c  c }
        \toprule
         \textbf{Level} & \textbf{SRCC} & \textbf{PCC} & \textbf{Tau-b} &\textbf{Tau-b 95} & \textbf{RMSE}\\
        
        \midrule
            {\small\textbf{Clip}} & 0.992  & 0.995  & 0.930 & 0.949  & 0.252 \\
            {\small\textbf{Model}} & 0.995 & 0.999  & 0.967 & 0.975 & 0.102\\
            
        \bottomrule
        \end{tabular}
    }
    \end{center}
\end{table}

\section{Experiments and Results}
\label{sec:results}
The FRVQA models predict the quality of a given compressed clip in comparison to a reference clip. This per clip score when compared to the DMOS per clip from the subjective test is referred to as \textit{clip level} evaluation. However, the main focus of this paper is to compare the codec performance, which is referred to as \textit{model level} evaluation, and accomplished by comparing the arithmetic mean of the DMOS values for all the videos encoded by that codec (i.e., DMOS per model), with the arithmetic mean of the predicted scores. It should be noted that each codec compressed the same set of videos, therefore the predicted scores are directly comparable. We evaluate the prediction of FRVQA models, at both the clip level and model level.

\begin{table*}[htbp]
    \caption{Performance of MLCVQA compared to other FRVQA models on predicting subjective DMOS, on the CLIC dataset (\ref{sec:clic_dataset}), in clip and model levels. Metrics are averaged over 5-fold cross-validation splits. Best values in each column (metric) are shown in bold font. Models that are statistically different from the winner at 95\% and 90\% confidence levels are indicated by superscripts \textbf{a} and \textbf{b} respectively.}
    
    \label{tab:results_5fold_final} 
    \centering
    \resizebox{0.8\textwidth}{!}{%
        \small
        \begin{tabular}{ l  c c c c    c   c c c c}
        \toprule
         \textbf{Objective}&	 \multicolumn{4}{c}{\textbf{Clip level}} & &  \multicolumn{4}{c}{\textbf{Model level}}\\
        
        \textbf{Model} & { \textbf{PCC$\uparrow$}} & { \textbf{SRCC$\uparrow$}}&	{\textbf{RMSE$\downarrow$}}&	{ \textbf{Tau-b 95$\uparrow$}} & &
        { \textbf{PCC$\uparrow$}} & { \textbf{SRCC$\uparrow$}}&	{\textbf{RMSE$\downarrow$}}& { \textbf{Tau-b 95$\uparrow$}} 
        \\ 
        \midrule
LPIPS             &  0.84{\textsuperscript{a}} &  0.88{\textsuperscript{a}} &   2.28{\textsuperscript{a}} &  0.71{\textsuperscript{a}} &&  0.95{\textsuperscript{a}} &  0.95{\textsuperscript{a}} &   2.38{\textsuperscript{a}} &  0.87{\textsuperscript{a}} \\
MS-SSIM           &  0.74{\textsuperscript{a}} &  0.87{\textsuperscript{a}} &   3.62{\textsuperscript{a}} &   0.72{\textsuperscript{a}} &&  0.89{\textsuperscript{a}} &  0.95{\textsuperscript{a}} &   3.60{\textsuperscript{a}} &  0.88{\textsuperscript{a}} \\
C3DVQA            &   0.90{\textsuperscript{a}} & 0.91{\textsuperscript{a}} &   1.28{\textsuperscript{a}} &  0.75{\textsuperscript{a}} &&  0.97{\textsuperscript{a}} &  0.97{\textsuperscript{b}} &   0.97{\textsuperscript{a}} &                        0.90 \\
PSNR              &  0.77{\textsuperscript{a}} &  0.77{\textsuperscript{a}} &   1.60{\textsuperscript{a}} &  0.60{\textsuperscript{a}} &&  0.92{\textsuperscript{a}} &  0.91{\textsuperscript{a}} &   1.18{\textsuperscript{a}} &  0.82{\textsuperscript{a}} \\
CompressedVQA     &    0.94{\textsuperscript{b}} &                       \textbf{0.94} &   1.01{\textsuperscript{a}} &  \textbf{0.81} &&  0.97{\textsuperscript{a}} &  0.96{\textsuperscript{a}} &   0.77{\textsuperscript{a}} &  0.89{\textsuperscript{a}} \\
VMAF (retrained)   &   \textbf{0.95} &    \textbf{0.94} &                        \textbf{0.85} &   \textbf{0.81} &&  0.98{\textsuperscript{a}} &  0.97{\textsuperscript{a}} &   0.49{\textsuperscript{a}} &   0.90{\textsuperscript{b}} \\
VMAF              &  0.92{\textsuperscript{a}} &  0.92{\textsuperscript{b}} &  0.96{\textsuperscript{b}} &  0.77 &&  0.96{\textsuperscript{a}} &  0.93{\textsuperscript{a}} &  0.71{\textsuperscript{a}} &  0.82{\textsuperscript{a}} \\
\midrule
MLCVQA (ours) &   0.94 &     \textbf{0.94} &    \textbf{0.85} &    0.80 &&                       \textbf{0.99} &       \textbf{0.99} &  \textbf{0.34} & \textbf{0.93} \\
\bottomrule
        \end{tabular}
    }
\end{table*}
\begin{table*}[tb]
\caption{Ablation study on different components of MLCVQA. Metrics are averaged over 5-fold cross-validation splits of the CLIC dataset \ref{sec:clic_dataset}. Best values in each column (metric) are shown in bold font. Results that are statistically different from the full model results at 95\% and 90\% confidence levels are indicated by superscripts \textbf{a} and \textbf{b} respectively.}
\centering
 \resizebox{0.8\textwidth}{!}{
\small
\begin{tabular}{cc cccc c cccc}
\toprule
\multirow{2}{*}{\textbf{Augmentation}} & \multirow{2}{*}{\begin{tabular}[c]{@{}c@{}}\textbf{Frame}\\ \textbf{Metrics}\end{tabular}} & \multicolumn{4}{c}{\textbf{Clip Level}} & & \multicolumn{4}{c}{\textbf{Model Level}} \\
&  & \textbf{PCC$\uparrow$} & \textbf{SRCC$\uparrow$} & \textbf{RMSE$\downarrow$} & \textbf{Tau-b 95$\uparrow$} & & \textbf{PCC$\uparrow$} & \textbf{SRCC$\uparrow$} & \textbf{RMSE$\downarrow$} & \textbf{Tau-b 95$\uparrow$} \\ 
\midrule
\checkmark & \checkmark & \textbf{0.94} & \textbf{0.94} & \textbf{0.85} & \textbf{0.80} & & \textbf{0.99} & \textbf{0.99} & \textbf{0.34 }&\textbf{ 0.93} \\ 
\midrule
\text{-} & \checkmark & 0.92 & 0.91 & 1.04 & 0.76 & &\textbf{0.99} & 0.98 & 0.41 & 0.92 \\
\checkmark & \text{-} & 0.92 & 0.91 & 0.99 & 0.78 & &\textbf{0.99} & 0.98 & 0.46{\textsuperscript{b}} & \textbf{0.93} \\
\text{-} & \text{-}  & 0.89 & 0.88 & 1.18{\textsuperscript{b}} & 0.74 & & \textbf{0.99}{\textsuperscript{b}} & 0.98 & 0.52{\textsuperscript{a}} & 0.92 \\ 
\bottomrule
\end{tabular}
 }

\label{tab:ablation_checkmarks}
\end{table*}
We report the performance of our model and other FRVQA models on the CLIC dataset (Section \ref{sec:clic_dataset}) with 5-fold cross-validation. For each fold, six videos (each encoded by 27 codecs) are chosen for the test split, while 24 videos are kept in the training split. For a fair comparison, we retrained our model, VMAF, C3DVQA, and CompressedVQA for each fold with and without augmented data and always report the best results for each method in Table \ref{tab:results_5fold_final}. For PSNR, MS-SSIM, VMAF, and LPIPS
\footnote{LPIPS calculates distance \textit{d} between a processed and a reference frame. For a video with $N$ frames, $1- \frac{1}{N}\Sigma_{i=1}^N d_i$ is considered as LPIPS' quality prediction.}
 we calculated their performance in each test fold as well. Performance of all FRVQA models\footnote{Best performing instance of each model (with or without augmentation) is reported here. See Section \ref{sec:augmentation}.}, averaged over 5 test folds are reported in Table~\ref{tab:results_5fold_final}. We perform a statistical significance test comparing the performance of each model metric with our model and marked cases with statistically significant differences. To the authors' best knowledge, this is the first VQA comparison that performs statistical significance tests.

VMAF, MLCVQA, and CompressedVQA performed closely in the clip level, where retrained VMAF and MLCVQA showed better accuracy (i.e., RMSE). In the model-level evaluation, MLCVQA performed better than all the other models in all metrics. 

Distribution of the predictions in both clip and model levels for MLCVQA, retrained VMAF, CompressVQA, LPIPS, MS-SSIM, and PSNR are illustrated in Fig.~\ref{fig:results:scatter}. MLCVQA shows excellent performance at the model level and competitive performance at the clip level. 
The theoretical maximum performance that can be achieved by an objective model given this test set was reported in Table~\ref{tab:max_expected_result}. Compared to that, MLCVQA's performance can be improved in clip-level for all metrics and in model-level in RMSE ($\delta=0.24$) and Tau-b 95 ($\delta=0.05$) metrics in future work.

\begin{figure}
    \centering
  \includegraphics[width = 0.9\columnwidth]{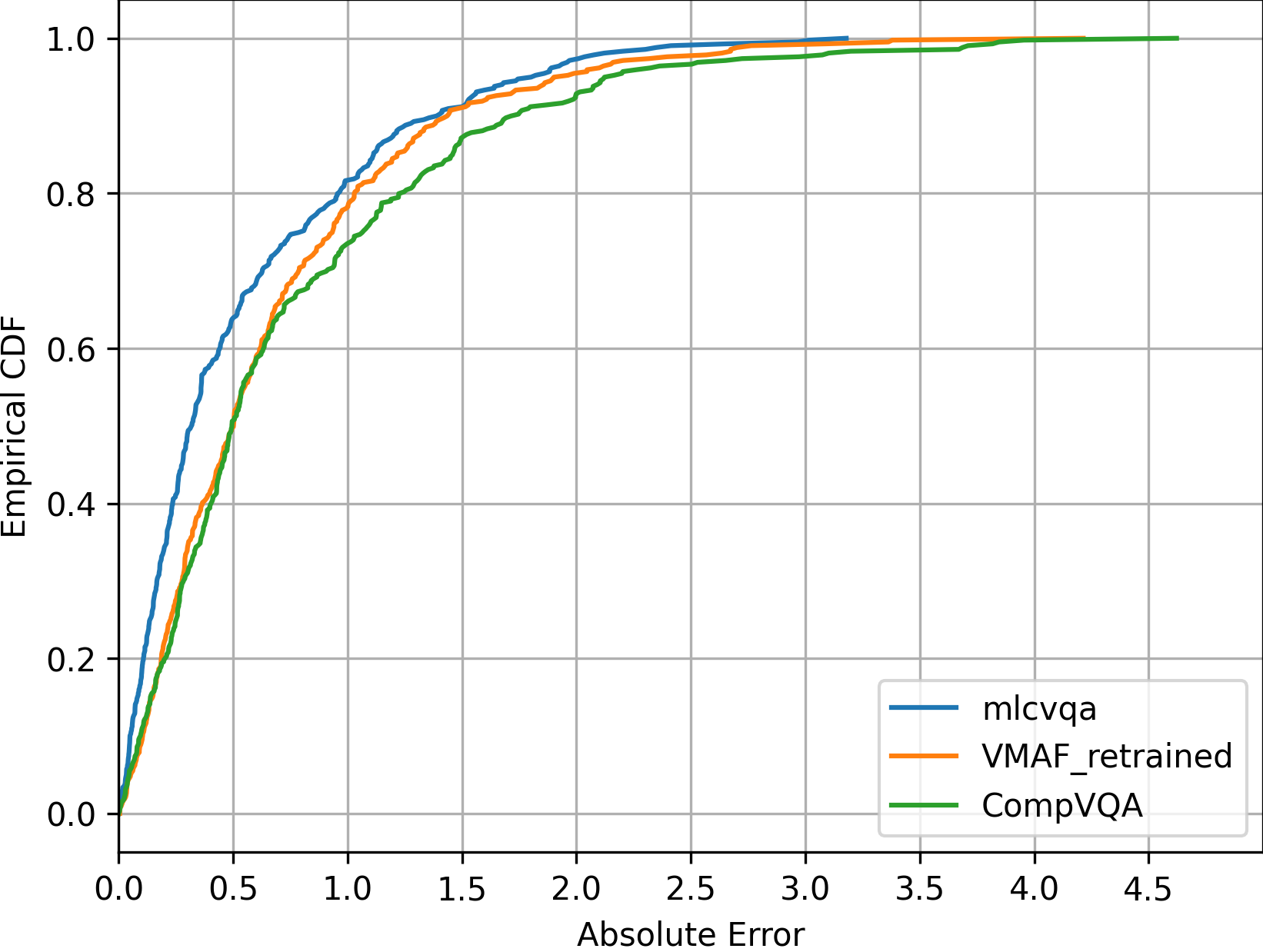}
  \caption{Empirical cumulative distribution of absolute error for the top-3 performing FRVQA models in the clip level.}
  \label{fig:cdf}
\end{figure}

Figure~\ref{fig:cdf} illustrates the empirical cumulative distribution of absolute errors. Although MLCVQA and the retrained VMAF obtained equal RMSE = 0.85, the prediction error of the MLCVQA was below 0.46 (i.e., the average 95\% CI of subjective ratings in clips level) more than 60\% of the time.

\subsection{Ablation Study}
In this section, we show how different components of MLCVQA contribute to its performance. Table \ref{tab:ablation_checkmarks} shows the results of our ablation study. Specifically, we test the contribution of (1) SlowFast features, (2) Augmentation, and (3) Adding image level metrics.  We observe how using SlowFast features alone results in very high performance at the model level. This performance is boosted by simple geometric augmentations and further improved by adding image-level metrics. 

\subsection{Performance with and without augmentation}
\label{sec:augmentation}
To utilize augmentation, we performed multiple experiments with different types, levels, and augmentation strategies. 
First,  within an expert viewing session, we established the highest level of augmentation without a noticeable difference in quality for all augmentations. Table \ref{tab:jnd_results} reports the type and level of augmentations. We also examined these levels in a follow-up DCR crowdsourcing test, where the quality of the augmented clips was rated compared to the reference clips. We observed less than a 1 DMOS change in quality for Rotation and Additive Gaussian Noise and less than a 0.5 DMOS change for the rest of the augmentations. Therefore, we considered these as minimum augmentation levels. 
\begin{table}[tb]
\caption{Augmentation types and their minimum level as a result of expert ratings.}
\centering
\resizebox{\columnwidth}{!}{
\begin{tabular}{ l l l l}
\toprule
{\textbf{Augmentation}} & {\textbf{Parameter Range}} & {\textbf{Baseline value}} & {\textbf{Value description}} \\
\midrule
\textbf{Additive Gaussian Noise} & [0, 0.025] & 0.0& \% of pixels affected \\ 
\textbf{Brightness} & [0.9, 1.1] & 1.0 & \% w.r.t baseline \\ 
\textbf{Center crop} & [0, 0.1] & 0.0 & \% w.r.t image shape \\ 
\textbf{Gamma} & [0.9, 1.1] & 1.0 & \%  w.r.t baseline \\ 
\textbf{Hue} & [-0.025, +0.025] & 1.0 & Value range \\ 
\textbf{Rotate} & [-1, 1] & 0.0 & Degree \\ 
\textbf{Saturation} & [0.75, 1.25] & 1.0 & \% w.r.t baseline \\ 
\bottomrule
\end{tabular}
 }

\label{tab:jnd_results}
\end{table} 
We examined different augmentation strategies with the following configurations: 1) only processed clips augmented with minimum levels (AugDistortedOnly), 2) Both reference and processed clips are augmented, with 3$\times$ of the minimum levels (ExtremeAug), 3) geometric augmentations as described in Section 4.2 of the paper and the subset of geometric augmentations from table \ref{tab:jnd_results} (GeometryOnly), 4) ensemble testing \cite{shanmugam_better_2021} where many augmentations are averaged for a final prediction value (TestAugmentation), 5) a subset of GeometryOnly where all geometric augmentations coming from Table~\ref{tab:jnd_results} were removed (SubsetGeometryOnly). We evaluated these strategies on the performance of the MLCVQA model. Results are reported in Table~\ref{tab:results_aug_5fold_final} and show that the SubsetGeometryOnly and TestAugmentation strategies beat the rest. We use SubsetGeometryOnly in the next steps as it is the simpler approach.

\begin{table*}[htp]
    \caption{Performance of ML-based models under different augmentation strategies. Metrics are averaged over 5-fold cross-validation splits. Models are compared to the winner (shown in bold font) within each group. There are three groups: first all augmentations are tested on MLCVQA\textsubscript{SF}, second, the winning augmentation strategy is applied to retrain C3DVQA and CompressedVQA, third, we add frame-level features to the best performing strategy to get MLCVQA (ours).
    Models that are statistically different from the winner within their group at 95\% and 90\% confidence levels are indicated by superscripts \textbf{a} and \textbf{b} respectively.  }
    
    \label{tab:results_aug_5fold_final} 
    \centering
    \resizebox{1\textwidth}{!}{%
        \small
        \begin{tabular}{ l l c c c c    c   c c c c}
        \toprule
         \textbf{Objective}&	\textbf{Augmentation}  & \multicolumn{4}{c}{\textbf{Clip level}} & &  \multicolumn{4}{c}{\textbf{Model level}}\\
        
        \textbf{Model} & \textbf{strategy} &  { \textbf{PCC$\uparrow$}} & { \textbf{SRCC$\uparrow$}}&	{\textbf{RMSE$\downarrow$}}&	{ \textbf{Tau-b 95$\uparrow$}} & &
        { \textbf{PCC$\uparrow$}} & { \textbf{SRCC$\uparrow$}}&	{\textbf{RMSE$\downarrow$}}& { \textbf{Tau-b 95$\uparrow$}} 
        \\ 
        \midrule

\multirow{4}{*}{{\small\textbf{MLCVQA\textsubscript{SF}\textsuperscript{1}}}}  & {\small\textbf{AugDistortedOnly}}  &  0.70{\textsuperscript{a}} &  0.73{\textsuperscript{b}} &  2.14{\textsuperscript{a}} &   0.60{\textsuperscript{b}} &&  0.96{\textsuperscript{a}} &  0.96{\textsuperscript{a}} &  1.88{\textsuperscript{a}} &  0.87{\textsuperscript{a}} \\
 &{\small\textbf{ExtremeAugmentations}}        &  0.87 &  0.86 &  1.23 &  0.71 &&  0.97{\textsuperscript{a}} &  0.96{\textsuperscript{a}} &  0.64 &  0.88{\textsuperscript{a}} \\
  &{\small\textbf{TestAugmentation}}  &                       0.92 &                       0.91 &                       0.99 &                       0.78 &&                       0.99 &                       0.98 &  0.46 &                       0.93 \\
  & {\small\textbf{GeometryOnly}}     &                       0.92 &                       0.91 &                       0.97 &                       0.77 &&  0.99 &  0.98 &                       0.43 &  0.92 \\
  & {\small\textbf{SubsetGeometryOnly}}     &    \textbf{0.92} &   \textbf{0.91} &   \textbf{0.99} & \textbf{0.78} &&  \textbf{0.99} &  \textbf{0.98} &  \textbf{0.46} &  \textbf{0.93} \\
\midrule
{\small\textbf{C3DVQA-aug}} &  
{{\small\textbf{SubsetGeometryOnly}}}   
&  0.89{\textsuperscript{a}} &  0.89{\textsuperscript{a}} &  1.44{\textsuperscript{a}} &  0.73{\textsuperscript{a}} &&  0.97{\textsuperscript{a}} &  0.96{\textsuperscript{a}} &  1.14{\textsuperscript{a}} &  0.89{\textsuperscript{a}} \\
{\small\textbf{CompressedVQA-aug}}& {{\small\textbf{SubsetGeometryOnly}}}    &  0.86{\textsuperscript{a}} &  0.89{\textsuperscript{a}} &  1.56{\textsuperscript{a}} &  0.72{\textsuperscript{a}} &&  0.96{\textsuperscript{a}} &  0.95{\textsuperscript{a}} &  1.17{\textsuperscript{a}} &  0.86{\textsuperscript{a}} \\
\midrule
{\small\textbf{MLCVQA (ours)}} & {{\small\textbf{SubsetGeometryOnly + Frame-level}}}     & \textbf{0.94} &    \textbf{0.94} &              \textbf{0.85} &  \textbf{0.80} &&   \textbf{0.99} &  \textbf{0.99} &              \textbf{0.34} & \textbf{0.93} \\
\bottomrule
\multicolumn{11}{l}{\small\textbf{\textsuperscript{1}} A  version of MLCVQA model only with SlowFast features.}\\
        \end{tabular}
    }
\end{table*}


We also evaluated the effect of augmentation on other ML-based FRVQA models. Table \ref{tab:results_aug_5fold_final} also shows the results of experiments with CompressedVQA and C3VQA, both retrained using SubsetGeometryOnly, the best set of performing augmentations for MLCVQA. The MLCVQA (ours) result adds frame-level features to the best-performing features which are described in Section 4.1 of the paper. 5-fold cross-validation results are used to compare all experiments with MLCVQA.
  
\subsection{MLCVQA performance and its limitations}

\begin{table}
        \caption{Average importance of top-performing features used in random forest regressors predicting the absolute error of MLCVQA.}
    
    \label{tab:rand_forestl} 
    \begin{center}
    \resizebox{0.7\columnwidth}{!}{%
        \begin{tabular}{l  c}
        \toprule         
        
         \textbf{Feature}&   \textbf{Avg. Importance} \\ 
        \midrule
            95\% CI & 0.366\\
            Codec & 0.109\\
            N of ratings & 0.082\\
            2.5\% Percentile of TI & 0.067\\
            Standard deviation of SI & 0.064 \\
            Minimum TI & 0.042 \\
            Bitrate & 0.037\\            
        \bottomrule
        \end{tabular}
    }
    \end{center}
\end{table}

We observed that MLCVQA is performing well in predicting the quality of videos on both extremes of the DMOS scale (1 and 9) and shows lower accuracy for videos in the middle-quality range. We trained a random forest regressor to predict the absolute error of MLCVQA prediction (per clip) given some features from source videos like SI and TI, and the uncertainty of subjective ratings. We randomly selected 80\% of the data for training and repeated the process 20 times with newly selected training and test sets. On average, the regressor's prediction on the test set has a PCC = 0.63 with the absolute error. The essential features and their average importance over all repetitions are reported in Table~\ref{tab:rand_forestl}. Results show that the uncertainty of subjective ratings is moderately correlated with the absolute error. However, it should be interpreted with caution as it is known that items located in the middle-quality range mostly have a larger 95\% CI compared to items located in both poles of the scale. There could be a confounding factor that both increases uncertainty in subjective rating and prediction error of MLCVQA. 

\begin{figure}
    \centering
  \includegraphics[width = 1\columnwidth]{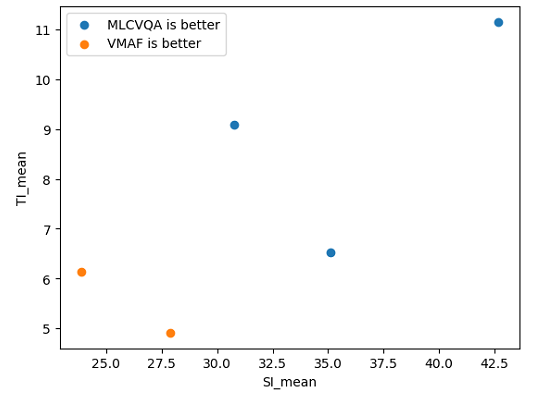}
  \caption{SI vs TI on different splits. Models are labeled as performing better if the SRCC is higher for that split.}
  \label{fig:TI-SI}
\end{figure}

Figure \ref{fig:TI-SI} shows the TI and SI for each split. In addition, we show if VMAF is performing better than MLCVQA based on which model has a higher SRCC for that split. Note that VMAF tends to do better for lower TI and lower SI values. 

\section{Conclusions}
\label{sec:conclusion}
We have provided the first large open-source dataset MLVC-FRVQA for ML video codec FRVQA development and an FRVQA model MLCVQA that exceeds the performance of the existing FRVQA methods that have tested. The MLVC-FRVQA dataset can be used to create a challenge to improve MLCVQA. MLCVQA can be used for ML video codec development for ranking new ML video codecs, hyper-parameter tuning, and training. 

There is still significant room for improvement for MLCVQA as measured using Kendall's Tau-b 95 and through the theoretical analysis done in Table \ref{tab:max_expected_result}. Future development includes utilizing the MLVC-FRVQA dataset for training. We can also try additional pre-trained models, including an ensemble of models that should improve MLCVQA's performance. We found that the leading predictor of the performance of MLCVQA for a specific clip is the 95\% confidence interval of the clip under evaluation. Therefore, we can significantly improve the results of MLCVQA by adaptively getting more quality ratings in the DMOS=[3,7] range where there is more uncertainty in the ratings (see Fig.~\ref{fig:results:scatter}), thereby reducing the confidence intervals for those clips. 

{\small
\bibliographystyle{ieee_fullname}
\bibliography{IC3-AI,Video_quality}
}

\end{document}


\title{Supplement to Full Reference Video Quality Assessment for Machine Learning-Based Video Codecs}

\author{First Author\\
Institution1\\
Institution1 address\\
{\tt\small firstauthor@i1.org}
\and
Second Author\\
Institution2\\
First line of institution2 address\\
{\tt\small secondauthor@i2.org}
}
\maketitle

\section{Introduction}
This document includes material for the submitted paper that we did not have room for in the main paper. We hope it clarifies some areas in the main paper and how we developed the MLCVQA model and MLVC-FRVQA dataset. 

\section{Supplement videos}
The supplement video ``CLIC examples sorted by DMOS.mp4'' shows an example of CVPR 2022 CLIC entries on one video clip processed at 0.1 Mbps (denoted *\_01) and at 1.0 Mbps (denoted *\_10), sorted by DMOS from worst to best quality. This video shows examples of the unique artifacts that ML video codecs have, such as ``wavy'' artifacts that are especially noticeable at 0.1 Mbps. The video ``CLIC examples single frame sorted by DMOS.mp4'' shows just a single frame for each CLIC entry, which makes it easier to see the artifacts. Compare the ML video codecs to libx264\_10 in particular, which is H.264 and has ``block'' artifacts typical of DSP codecs.

{\small
\bibliographystyle{spconf}
\bibliography{IC3-AI}
}